\documentclass{aa}  
\usepackage{natbib}
\usepackage{graphicx}
\usepackage{txfonts}
\newcommand{\teff}{$T_{\rm eff}$}
\newcommand{\kms}{km\,s$^{-1}$}
\newcommand{\lgg}{$\log{g}$}
\newcommand{\vsini}{$v\sin{i}$}
\newcommand{\vs}{$v_{\rm e}\sin i$}
\def\bs{$\langle B_{\rm s} \rangle$}

\def\br{$\langle B_{\rm r} \rangle$}

\newcommand{\bmag}{\textsc{BinMag6}}
\newcommand{\vald}{\textsc{VALD}}

\def\ione{\,{\sc i}}
\def\ii{\,{\sc ii}}
\def\iii{\,{\sc iii}}

\begin{document}

   \title{Fundamental parameters of the Ap-stars GO And, 84 UMa, and $\kappa$~Psc.}

   \author{A. M. Romanovskaya\fnmsep\thanks{annarom@inasan.ru}
          \inst{1}
         \and
          D. V. Shulyak\inst{2}      
          \and
          T. A. Ryabchikova\inst{1}
          \and
          T. M. Sitnova\inst{1}
  }

   \institute{Institute of Astronomy, Russian Academy of Sciences, Pyatnitskaya 48, 119017, Moscow, Russia\\
         \and
             Instituto de Astrof\'{\i}sica de Andaluc\'{\i}a - CSIC, c/ Glorieta de la Astronom\'{\i}a s/n, 18008 Granada, Spain.\\
             }
   \date{Received April 00, 000; accepted March 00, 0000}

% \abstract{}{}{}{}{} 
% 5 {} token are mandatory
 
  \abstract
  %{} leave it empty if necessary  
   %{Physical processes in magnetic stars due to the presence of magnetic fields complicate the fundamental parameters determination. At each stage of the study, it is necessary to carefully consider the presence of a magnetic field}
   {}
   {The aim of this work is to determine fundamental parameters of three Ap stars, GO And (HD~4778), $\kappa$~Psc (HD~220825), and 84 UMa (HD~120198), using
   spectroscopic techniques. By analysing these stars, we complete the sample of Ap stars for which fundamental parameters have additionally been derived by means of interferometry. This enables a cross-comparison of results derived by direct and indirect methods.} 
   {Our study is based on the analysis of high-resolution spectra with a high signal-to-noise ratio that were obtained with ESPaDOnS spectrograph. We used an iterative method of fundamental parameter determinations that includes self-consistent modelling of the stellar atmosphere, taking individual abundances of chemical elements into account and subsequently fitting a theoretical spectral energy distribution to the observed distribution. The quality of the spectroscopic determinations was evaluated through a comparison with the interferometric results.}
   {For all investigated stars, we determined fundamental parameters and derived chemical abundances that are  typical for Ap stars. The abundances are mainly characterised by a gradual increase of heavy element atmospheric abundances from an order of magnitude for iron peak elements  up to very significant excesses of 3--4~dex of the rare-earth elements relative to the solar values. The only exception is Ba, whose abundance is close to the solar abundance. There is also a significant He deficiency in the atmospheres of HD~120198 and HD~220825, whereas the He abundance in HD~4778 is close to the solar abundance. We do not find significant Fe and Cr stratification. Using these abundances, we constructed self-consistent atmospheric models for each star. The effect of the surface chemical inhomogeneity on the derived fundamental parameters did not exceed $\pm$100~K in effective temperature, which lies within the range of errors in similar self-consistent analyses of Ap stars. Finally, we compared spectroscopically derived effective temperatures, radii, and luminosity for 13 out of 14 Ap stars in a benchmark sample with the interferometric results. While radii and luminosity agree within the quoted errors of both determinations,  the spectroscopic effective temperatures are higher than the interferometric temperatures for stars with \teff$>$9000~K. The observed hydrogen line profiles favour the spectroscopically derived temperatures.}
   {}
   

   \maketitle
%
%-------------------------------------------------------------------

\section{Introduction}

Magnetic chemically peculiar (Ap) stars belong to the main-sequence (MS) group between spectral types B5 to F5. These stars have global magnetic fields in which the measured longitudinal components vary from a few dozen Gauss \citep{2007A&A...475.1053A} to tens of kilo-Gauss \citep[see the catalogue of][]{2008AstBu..63..139R}. The magnetic fields have a predominantly poloidal structure. These stars are also slow rotators with rotation velocities lower than 100~\kms,  which is thought to be due to magnetic braking \citep{2017A&A...601A..14M}. Ap stars have the same masses and luminosity as normal MS stars with similar temperatures.

Strong magnetic fields significantly affect the atmospheric chemical composition in Ap-stars. The observed abundance anomalies were explained by Michaud \citep{1970ApJ...160..641M} to be a result of particle diffusion. Studies of the stratification of chemical elements by spectral observations showed that most of the elements have vertical abundance gradients \citep{2009A&A...499..879S}.

For Ap stars, the standard photometric and spectroscopic calibrations that were developed to determine the fundamental parameters of normal stars are often inapplicable. In magnetic stars, the presence of magnetic fields and significant individual chemical composition anomalies requires a detailed study of stellar chemistry in order to construct an adequate model atmosphere that can predict the observed flux distribution. A self-consistent procedure for the simultaneous modelling of the spectrum and flux distribution was first proposed by 
\citet{2009A&A...499..879S} for the Ap star HD~24712. Then, this method was used to determine the fundamental parameters of several other Ap stars \citep{2009A&A...499..851K, 2010A&A...520A..88S, 2011MNRAS.417..444P, 2013A&A...551A..14S, 2013A&A...552A..28N}. 

The accuracy of determining fundamental parameters by spectroscopic methods can be estimated by comparing them with the results of direct determinations of stellar radii  by means of interferometric observations. Combining the latter with the (spectro)photometric absolute calibrated flux, stellar temperature and luminosity
can be estimated for a known distance. Interferometry provides a model-independent way to estimate stellar radii that can serve as a reference for testing theoretical models. 
Recently, \citet{2020A&A...642A.101P} presented interferometric observations of a sample of Ap stars
obtained with CHARA array\footnote{http://www.chara.gsu.edu/} \citep{2010SPIE.7734E..03T} that started in 2008 (see their Table~1). The sample contains stars of late-B~to~early-F spectral types and diverse stellar ages so that the atmospheric properties of these stars such as magnetic field strength and chemical abundances can substantially differ from star to star. The sample was primarily chosen according to instrument limitations, which required stars to be bright (V$^m <$6) and have relatively high angular diameters $>0.2$~mas. Given an unique opportunity to test our theoretical models and analysis methods against independent approaches, we initiated a follow-up study of interferometric targets by means of spectroscopy. Our early results indicated a good agreement between two methods \citep[see e.g.][]{2019MNRAS.488.2343R}. The aim of this paper 
is to complete the follow-up spectroscopic program
and present the results of a detailed spectroscopic analysis of the last three Ap stars from the sample of interferometric measurements: HD~4778 (GO And), HD~120198 (84 UMa), and HD~220825 ($\kappa$ Psc). We also present a final comparison between the results obtained by direct (interferometry) and indirect (spectroscopy) methods.

%--------------------------------------------------------------------
\section{Observations}
\label{obs}
Spectroscopic observations were extracted from the ESPaDOnS spectrograph (Canada-France-Hawaii Telescope) archive\footnote{http://www.cadc-ccda.hia-iha.nrc-cnrc.gc.ca/en/cfht/} and cover the spectral range of 3700-10\,500\AA. The resolving power of the instrument is $R$$=$$\lambda/\Delta\lambda$$=$$68\,000$. Spectra of these stars normalised to the continuum level are provided as supplementary material in the form of an ASCII file. This is available in electronic form at the CDS\footnote{http://cdsweb.u-strasbg.fr/cgi-bin/CatS/2.92.198.28}.

Photometric and spectrophotometric observations in different spectral regions were employed to construct the spectral energy distribution (SED) for each star. For the UV region, we used observations obtained with the S2/68 telescope of TD1 mission (European Space Research Organization (ESRO) satellite), which measured  the absolute flux in four bands, 1565\AA, 1965\AA, 2365\AA, and 2740\AA\, \citep{1978csuf.book.....T},  and data from the International Ultraviolet Explorer (IUE)\footnote{http://archive.stsci.edu/iue/}. In the optical range, we used data from Adelman's spectrophotometric catalogue \citep{1989A&AS...81..221A} and/or Johnson photometry in V filter taken from the ADS SIMBAD online database\footnote{http://simbad.u-strasbg.fr/simbad/}. Photometric data in the IR region were taken from the 2Micron All-Sky Survey (2MASS; \citet{2003yCat.2246....0C}), which contains an overview of the entire sky in J (1.25 microns), H (1.65 microns), and Ks (2.17 microns) filters. Observations were transformed into absolute flux using the calibrations given in \citet{2003AJ....126.1090C}.

For each of our sample stars we constructed a set of observed fluxes using the sources mentioned above. For HD~4778 we used the UV data from the IUE and TD1 missions. The optical range lacks Adelman's spectrophotometry, and thus we used only Johnson photometry, as well as 2MASS photometric measurements in the IR. For HD~120198 the photometric and spectrophotometric data were taken from the IUE, TD1, Adelman's, and 2MASS catalogues. In case of HD~220825 the IUE and TD1 data are lacking, and thus we used only Adelman's and 2MASS measurements to build the SED.

%--------------------------------------------------------------------

\section{Spectroscopic analysis}
\label{Abund}

\subsection{Methods}\label{Method}

The fundamental parameters of the stars were obtained by various methods in previous works (Table~\ref{literat-param}). Here and throughout the paper, an error in the measurement of the last digits is given in parentheses. The latest parameter determinations were performed by interferometry \citep{2020A&A...642A.101P}. \citet{2006A&A...450..763K} estimated the effective temperatures and luminosity of stars through the calibrations of the Geneva photometric system using the revised HIPPARCOS parallaxes \citep{2007A&A...474..653V}. The stellar radii were calculated using the Stefan-Boltzmann formula. \citet{2007AN....328..475H} determined the effective temperatures from the Geneva system and from Str\"{o}mgren system photometry, and the luminosity was obtained by taking the bolometric corrections into account. \citet{2008MNRAS.385..481L} determined \teff\, from fitting theoretical flux to the observed SED; \lgg\, in this work  was assumed to be \lgg\, = 4.0. \citet{2000A&A...355..308A} derived fundamental parameters through different methods. \citet{2008A&A...491..545N} collected fundamental parameters from different sources; the values of \teff\, given in Table~\ref{literat-param} were obtained using the IR flux method.

Here, we used an iterative method to determine stellar fundamental parameters \citep[for more details, see][]{2009A&A...499..879S}. It includes abundance and stratification calculations, if necessary, and  the subsequent refinement of the atmospheric model by fitting the calculated flux to the observed SED, as well as comparing observed and predicted hydrogen line profiles. This means that the abundance analysis and parameter determinations are interconnected. This allows us to obtain fairly accurate atmospheric parameters, normally within a few iterations.
 
 \begin{table*}[t]
        \label{refs}
        \caption{Literature data for the fundamental parameters of the program stars.}
        \small
        \centering
        \begin{tabular}{r l|l|l|l|l|c}
        \hline \hline HD  & \teff & \lgg & $\log(L/L_{\odot})$ & $R/R_{\odot}$ & Ref. & Method\\
        \hline
        4778      & ~9977(300) &~~~~~~& 1.51(07) & 1.91(04) & {\citet{2006A&A...450..763K}} & Photometry \\
                  & ~9135(400) &~~~~~~& 1.54(05) & 2.36(12) & {\citet{2020A&A...642A.101P}}& Interferometry\\
                  & ~9375(300) & 4.14(09) & 1.48(07) & 2.10(22) & {\citet{2007AN....328..475H}}& Photometry \\
        \hline
        120198 & 10543(390) &~~~~~~& 1.65(06) & 2.01(01) & {\citet{2006A&A...450..763K}} & Photometry \\%(250)
                  & ~9865(370) &~~~~~~& 1.65(04) & 2.28(10) & {\citet{2020A&A...642A.101P}}& Interferometry\\
                  & ~9750 & 4.00 &~~~~~~~~~~& 2.22 & {\citet{2008MNRAS.385..481L}}& SED\\
                          & 10500 & 4.19 &~~~~~~~~~~&~~~~~~& {\citet{2000A&A...355..308A}} & Spectrophotometry\\
                          & 11234 & 4.07 &~~~~~~~~~~&~~~~~~& {\citet{2000A&A...355..308A}} & $ubvy\beta$\\
                          & 10400 & 4.00 &~~~~~~~~~~&~~~~~~& {\citet{2000A&A...355..308A}} & $H_{\gamma}$\\
        \hline    
        220825 & ~9078(290) &~~~~~~& 1.35(04) & 1.92(03) & {\citet{2006A&A...450..763K}} & Photometry \\%(200)
                          & ~8790(230) &~~~~~~& 1.24(04) & 1.78(03) & {\citet{2020A&A...642A.101P}}& Interferometry\\
                      & ~9200(80)  & & & & {\citet{2008A&A...491..545N}}& IRFM \\
                          & ~9180(370)  & 4.27(07)  & 1.28(03)   &  1.73(14) & {\citet{2019MNRAS.483.2300S}} & SED \\
                      & ~9080(290)  & 4.16(08)  & 1.383(53)  & 2.00(18) & {\citet{2007AN....328..475H}} & Photometry \\
                  & ~9250 & 3.75 & & & {\citet{1996AstL...22..822R}} & IRFM \\
                \hline 
        \end{tabular}   
        \tablefoot{Several studies do not provide measurement uncertainties.}
\label{literat-param}
\end{table*}
 
To calculate the model atmosphere, it is necessary to know the basic parameters of the star, that is, the effective temperature \teff, the logarithm of the surface gravity \lgg, the metallicity, and also the magnetic field strength. We took initial atmospheric parameters from the work by \citet{2006A&A...450..763K}. The initial model atmosphere were calculated using the ATLAS9 program \citep{1993KurCD..13.....K}. In this model, the absorption in spectral lines is taken into account statistically through the opacity distribution function (ODF) as part of the overall opacity. The ATLAS9 model was used for a preliminary abundance analysis in chemically homogeneous stellar atmospheres. These calculations were carried out using the spectral synthesis program {\sc Synmast} \citep[for a detailed description, see][]{2007pms..conf..109K, 2010A&A...524A...5K}, which  takes the magnetic field in the radiative transfer calculation into account. The atomic parameters of spectral lines were taken from the third version of the Vienna Atomic Line Database \vald\, \citep{1999A&AS..138..119K, 2015PhyS...90e4005R}.

Atmospheric parameters were refined by comparing the observed and theoretical SEDs calibrated to absolute flux units. For this and the following steps of the iterative procedure, we used the code \textsc{LLmodels}  for the model atmosphere calculations \citep{2004A&A...428..993S}, which accounts for individual atmospheric abundances in opacity calculation. 

\subsubsection{Magnetic field and rotation velocity}

To determine the element abundance, we calculated the synthetic spectrum and fitted it to the observed spectrum using the visualisation program \bmag\, \citep{2018ascl.soft05015K}. The magnetic fields of our program stars are strong enough to play a significant role in the formation of spectral lines and therefore should be taken into account in the spectrum synthesis. Because the rotation velocity \vsini\ is rather high, it was not possible to estimate the surface magnetic field modulus \bs\ from Zeeman splitting  or from magnetic differential broadening \citep{2020AstL...46..331R} for all stars. Therefore we took \bs\ values from the literature:  2.6 kG for HD~4778  \citep{2019AstBu..74...66G}, 1.6 kG for HD~120198, and 2.0 kG for HD~220825 \citep{2017A&A...597A..58K}. These values of the magnetic field were employed in all synthetic spectrum calculations.  

We derived \vs\ from several line blends with different sensitivity to the magnetic field. To do this, we used the fitting procedure implemented in the \bmag\ code. The final values are the following: \vs = 31.6$\pm$3.8~\kms\ (HD~4778), \vs = 56.0$\pm$0.5~\kms\ (HD~120198), and \vs = 38.4$\pm$1.5~\kms\ (HD~220825).

\subsubsection{Chemical abundances}\label{Abun}

The average abundance of the chemical elements was determined with two methods. First, 
we determined the abundance through the measured equivalent widths of the spectral lines using the  code \textsc{WidSyn} \citep{2013A&A...551A..14S}.  \textsc{WidSyn} calculates theoretical equivalent widths by taking polarisation in spectral lines into account. We assumed the surface magnetic field to be constant with atmospheric depth and defined by the radial component of the magnetic field vector \br. This method is fast, but was not always useful because the rotation velocity \vs\ of stars is rather high and the line spectrum is rich, which causes the lines of different elements to blend and affects measurements of their equivalent widths. In order to obtain more accurate values, we therefore used the method of fitting the synthetic line profiles to the observed profiles. In this method, the following parameters must be optimised: the rotation velocity along the line of sight \vs, the radial velocity $V_r$, the radial component of the magnetic field $B_r$  and if necessary, the meridional $B_m$ component of the magnetic field, and finally, the element abundance. We fit the theoretical line profiles to the observed profiles using the code \bmag\, along with the code {\sc Synmast}  \citep{2007pms..conf..109K}. Here and below, the abundance is given as the logarithm of the ratio of the number of given element atoms to the total number of atoms of all elements $\log(N_{el}/N_{tot})$. 
Examples of a comparison between the two methods of the abundance determination are presented in Table~\ref{hd220825-abunds-ew-sp} for HD~220825. The equivalent width method
results in systematically higher abundances compared to the direct line profile fitting because the line profile is distorted: the lines of different elements blend with the spectral line of an element whose abundance we wish to measure. This artificially increases the width of the spectral line under investigation, and a higher abundance is needed to reproduce the measured equivalent width. In contrast, the direct spectrum synthesis self-consistently accounts for line blending and distortion of the shapes of spectral lines. Therefore we finally adopted the abundance values derived using spectrum synthesis.

\begin{table}[hbt!]
        \caption{Comparison of several elemental abundances in I and II ionisation stages, calculated through equivalent widths and spectrum synthesis for HD~220825.}
        \label{hd220825-abunds-ew-sp}
        \centering
        \begin{tabular}{l | c c}
                \hline  
                Ion & \(log\left(\frac{N}{N_{tot}}\right)_{EW}\) & \(log\left(\frac{N}{N_{tot}}\right)_{SS}\) \\
                \hline
                Mg\ione& -3.64(24) & -3.92(23) \\
                Mg\ii  & -3.84:    & -4.45:    \\
                Fe\ione& -3.23(40) & -3.37(23) \\
                Fe\ii  & -3.08(34) & -3.58(30) \\ 
                Mn\ii  & -4.38(25) & -4.82(05) \\
                Ti\ii  & -5.03(31) & -5.91(31) \\
                Cr\ione& -3.57(36) & -3.79(24) \\ 
                Cr\ii  & -3.55(25) & -3.84(36) \\
                Nd\iii & -6.66(61) & -6.82(25) \\
                Eu\ii  & -7.51(61) & -8.38(03) \\
                Eu\iii & -6.37:    & -6.40:    \\
                \hline
        \end{tabular}
\end{table}

For elements up to Ba, the abundances were obtained from the lines of neutral and first ions, while for the rare-earth elements (REE), we used the lines of the first and second ions. For all elements but O, the abundances were obtained in the local thermodynamic equilibrium (LTE) approximation. The oxygen abundance was determined from the lines of the IR triplet O\ione\, 7771--7775 \AA, taking the deviation from LTE (NLTE) into account. We adopted the O\ione\ model atom of \citet{Przybilla2000}, which was updated by including accurate data for collisions with electrons and hydrogen atoms as described in \citet{2013AstL...39..126S} and \citet{2018AstL...44..411S}, respectively.
The isotopic structure of Ba\ii\ and Eu\ii\ was taken into account, but we neglected hyperfine splitting (HFS) due to the complex interaction between hyperfine and magnetic splitting \citep{1975A&A....45..269L}. Therefore we caution that the abundance of elements with odd isotopes (Mn, Pr, Eu) might be overestimated.

\begin{table}
        \caption{Mean values of chemical abundances $log(N_{el}/N_{tot})$, calculated from equivalents widths and spectrum synthesis of N lines for model atmospheres 9605g40 (HD~4778), 10174g42 (HD~120198), and 9470g42 (HD~220825).} %The * marks the ions, the abundances of which was determined by the spectrum synthesis method.}
        \scriptsize
        \centering
        \begin{tabular}{l|cr|cr|cr|c}
                \hline \hline Ion & \multicolumn{7}{c}{Abundance,  $log(N_{el}/N_{tot})$} \\ 
                \hline & HD~4778 & N & HD~120198 &  N &  HD~220825 & N & Sun\\
                \hline 
                C\ione  & -3.80(50) & 1 & -3.31(50) & 3 & -4.26(15) & 2 & -3.61\\
                N\ione  & -4.63(50) & 1 &~~~~~~~~~~~&~~~& -5.02(50) & 1 & -4.21\\ 
                O\ione  & -4.84(08) & 3 & -5.21(07) & 3 & -5.33(07) & 3 & -3.35\\ 
                Na\ione & -5.91(50) & 1 &~~~~~~~~~~~&~~~& -4.81(50) & 1 & -5.80\\ 
                Mg\ione & -4.77(56) & 4 & -4.37(41) & 3 & -3.92(23) & 4 & -4.44\\ 
                Mg\ii   & -4.78(28) & 3 & -4.19(24) & 2 & -4.45(50) & 1 & \\ 
                Al\ione &~~~~~~~~~~~&~~~&~~~~~~~~~~~&~~~& -4.14(50) & 1 & -5.59\\
                Al\ii   & -4.76(29) & 2 & -5.19(01) & 2 & -4.67(50) & 1 & \\ 
                Si\ione &~~~~~~~~~~~&~~~& -3.84(10) & 2 & -4.19(50) & 1 & -4.53\\ 
                Si\ii   & -4.67(17) & 3 & -4.08(12) & 3 & -5.72(38) & 3 & \\ 
                Ca\ione & -4.08(01) & 2 & -5.60(50) & 1 & -4.99(50) & 1 & -5.70\\ 
                Ca\ii   & -4.80(20) & 2 & -6.37(07) & 2 & -5.71(50) & 1 & \\ 
                Ti\ii   & -5.95(25) & 5 & -6.34(54) & 4 & -5.91(31) & 7 & -7.09\\
                Cr\ione & -3.73(36) & 11 & -3.38(16) & 10 & -3.79(24) & 11 & -6.40\\ 
                Cr\ii   & -3.75(34) & 24 & -3.36(20) & 11 & -3.84(36) & 31 & \\ 
                Mn\ione & -4.95(50) & 1 &~~~~~~~~~~~&~~~~& -5.36(50) & 1  & -6.61\\ 
                Mn\ii   & -4.56(29) & 11 & -5.75(50) & 1 & -4.82(05) & 2  &  \\ 
                Fe\ione & -2.98(27) & 16 & -3.26(23) & 5  & -3.37(23) & 22 & -4.54\\ 
                Fe\ii   & -3.04(25) & 38 & -3.36(25) & 14 & -3.58(30) & 33 & \\ 
                Sr\ione & -5.57(50) & 1 &~~~~~~~~~~~&~~~& -5.47(50) & 1 & -9.17\\ 
                Sr\ii   & -7.10(50) & 1 & -7.51(50) & 1 & -7.54(50) & 1 & \\  
                Y\ii    & -7.61(50) & 1 &~~~~~~~~~~~&~~~&~~~~~~~~~~~&   & -9.83\\ 
                Zr\ii   & -8.53(50) & 1 &~~~~~~~~~~~&~~~&~~~~~~~~~~~&   & -9.46\\
                Ba\ii   & -9.82(50) & 1 & -9.50(50) & 1 & -8.41(50) & 1 & -9.86\\
                La\ii   & -7.35(13) & 2 & -8.17(50) & 1 & -8.50(50) & 1 & -10.94\\
                Ce\ii   & -6.99(31) & 7 & -7.29(50) & 1 & -7.45(50) & 1 & -10.46\\ 
                Pr\iii  & -7.86(26) & 6 & -8.34(05) & 2 & -7.26(23) & 3 & -11.32\\ 
                Nd\ii   & -7.72(50) & 1 &~~~~~~~~~~~&~~~& -8.20(50) & 1 & -10.62\\ 
                Nd\iii  & -7.56(35) & 7 & -7.94(43) & 9 & -6.82(25) & 5 & \\ 
                Sm\ii  & -7.63(10) & 3 & -7.75(50) & 1 & -7.80(50) & 1 & -11.08\\
                Eu\ii   & -7.31(16) & 4 & -7.65(41) & 4 & -8.38(03) & 2 & -11.52\\ 
                Eu\iii  & -5.47(50) & 1 & -6.60(50) & 1 & -6.37(50) & 1 & \\
                Gd\ii   & -8.22(50) & 1 &~~~~~~~~~~~&~~~&~~~~~~~~~~~&   & -10.97\\
                Lu\ii   & -8.47(30) & 2 &~~~~~~~~~~~&~~~& -7.77(50) & 1 & -11.94\\
                \hline 
        \end{tabular}
        \tablefoot{Values derived from single lines are marked with a colon. The standard deviation is given in parentheses, and was assumed to be 0.5~dex when the measurement was obtained from a single line only. The last column contains solar photospheric abundances that were taken from \citet{2021A&A...653A.141A}.
        }
        \label{MeanAbunds}
\end{table}

\begin{figure}[hbt!]
        \centering
        \includegraphics[width=0.70\linewidth, angle=90, clip]{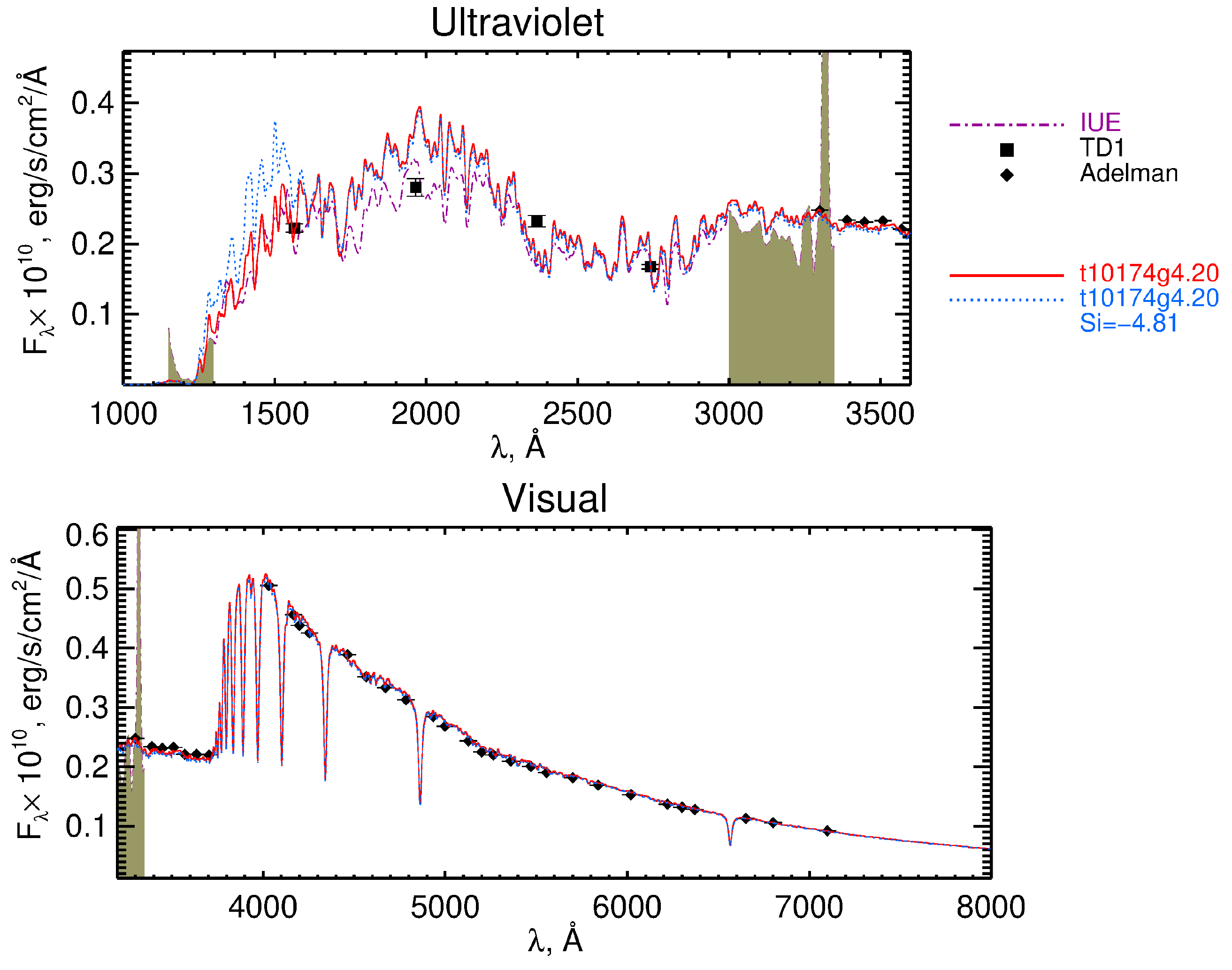}
        \caption{Spectral energy distribution of HD~120198 in the UV and visual spectral regions. Red and dotted blue lines show theoretical SED calculations with the same final model atmosphere parameters t10174g42, but with a different Si abundance: $log(N_{Si}/N_{tot})=-3.82$ and -4.81, respectively. Shaded parts indicate spectral regions excluded from the fit.}\label{HD120198-sed-Si}
\end{figure}

For HD~120198, it was impossible to derive an accurate Si abundance from the spectrum synthesis due to a lack of suitable unblended lines. However, we realised
that a change in the Si abundance noticeably affects the part of the flux in the UV spectral region (see Fig.~\ref{HD120198-sed-Si}). Decreasing the Si abundance by 1.0~dex leads to an overestimated flux in the region of 1500\AA\, compared to the observed data from the IUE and TD1. At the same time, in other regions of the spectrum, the fit remained unchanged, with only a slight increase in the effective temperature by about 50~K. This means that the flux observations in the UV spectral region may serve as an important indicator of the Si abundance in the atmospheres of some Ap stars, similar to what was observed in hotter B-type chemically peculiar stars \citep[see e.g.][]{2012A&A...537A..14K}.

\begin{figure*}[hbt!]
\centering
         \includegraphics[width=0.9\textwidth, clip]{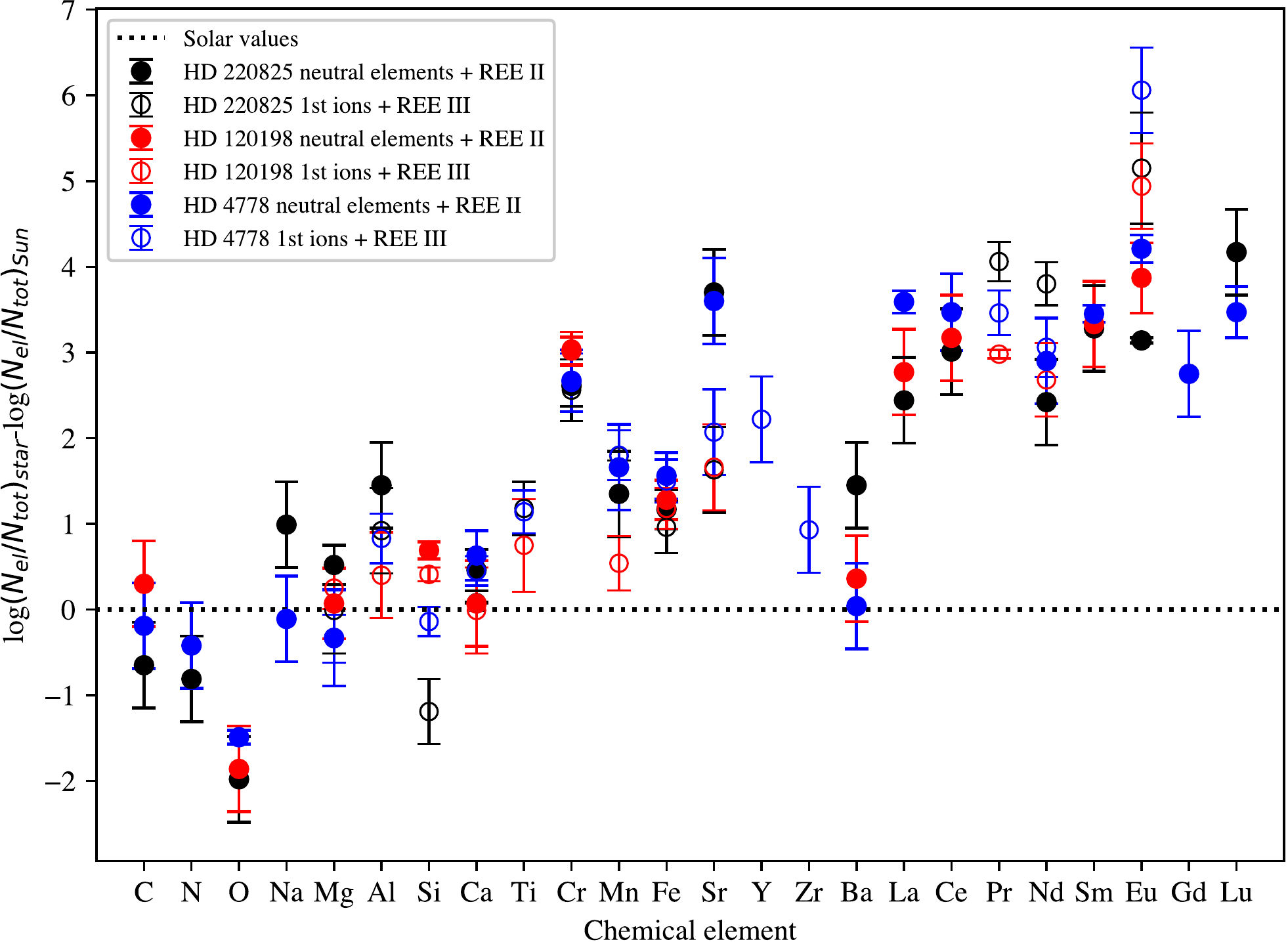}
        \caption{Atmospheric abundances in HD~4778 (blue circles), HD~120198 (red circles), and HD~220825 (black circles) relative to the solar values.}\label{abund}
\end{figure*}

 The final abundances in the atmospheres of the program stars together with the number of lines used in the analysis are collected in Table~\ref{MeanAbunds} and are shown in Fig.~\ref{abund} relative to the solar values.
 
  In summary, the abundances were estimated for 24 elements. For 7 elements, the abundances were derived from the  lines of two consecutive ions. In all three stars, we obtained typical abundance trends observed for Ap stars of SrCrEu type: a deficiency of C, N, and O; a close to solar abundance of Na, Mg, and Ca; a large Cr, Sr, and REE overabundance in agreement with previous works (see e.g. \cite{2010A&A...520A..88S, 2019MNRAS.488.2343R}).  A deficit of He is observed in stars HD~120198 and HD~220825 because the He\ione\ 4471~\AA\ line is practically not visible in their spectra. Therefore He-weak atmosphere models were calculated with a He abundance of log(He/N$_{tot}$))=-4.05 and, accordingly, with an H/N$_{tot}$ = 0.99. A similar He deficiency was observed for another Ap star, HD~108662 \citep{2020INASR...5..219R}. In the atmosphere of star HD~4778, He\ione\ is observed, and we used the solar abundance log(He/N$_{tot}$))=-1.05 and H/N$_{tot}$ = -0.91. 
  
 The ionisation equilibrium is conserved within the error bars  for Mg, Ca, Cr, Mn, and Fe. This shows the absence of significant vertical stratification of these elements. We consider the possible effect of Fe stratification on flux distribution in Section~\ref{strat}. The only REE with observable lines of two ions (first and second) is Eu, which shows the typical REE-anomaly that is observed in most Ap stars in the effective temperature range of 7000~K~--~10\,000~K \citep{2017AstL...43..252R}. The most interesting feature in the abundance distribution is the abundance sequence Sr-Y-Zr-Ba-REE, where a sharp decrease in the Ba abundance up to the solar value is observed and a $\approx$3~dex overabundance of the neighbouring elements Sr and REEs. Overabundances (to a lesser extent) of the heavy elements Sr-Zr-Ba-Nd were also derived in normal A stars and in Am stars \citep{2020MNRAS.499.3706M}, but Ba had the highest overabundance of the neighbouring elements Sr, Zr, and Nd there. This different behaviour of Ba certainly indicates a difference in the mechanisms that control the atmospheric abundance distribution in normal and peculiar A-type stars.

\subsubsection{Stratification analysis}\label{strat}

The vertical distribution of chemical elements in the stellar atmosphere (stratification profile) depends on the effective temperature
\teff, surface gravity \lgg,\, and magnetic field strength \bs\ of the star \citep{2009A&A...495..937L, 2010A&A...516A..53A}. Theoretical diffusion calculations of some elements provided by \citet{1992A&A...258..449B} and later by \citet{2009A&A...495..937L} showed that as a first approximation, the stratification profile of an element can be represented by a step function with four parameters: the element abundance in the upper atmosphere, the element abundance in the lower atmosphere, the location of the abundance jump, and the width of this jump. This was then widely used in the analysis of spectral observations (e.g. \citet{2001ASPC..248..373W, 2005A&A...438..973R}).

The accuracy of the stratification study strongly depends on the choice of spectral lines. To study the stratification in the atmospheres of our sample stars, a set of the most suitable lines must be chosen. This includes individual and preferably unblended lines with different excitation potentials $E_{i}$, different intensities and equivalent widths, that is, lines that are formed at different optical depths and thus probe different atmospheric layers. In the atmospheres of Ap stars, the elements Fe and Cr are the best elements for the stratification analysis because a large number of spectral lines are available. 

\begin{figure*}[hbt!]
        \begin{minipage}[h]{0.40\linewidth}
                \center{\includegraphics[width=1\linewidth]{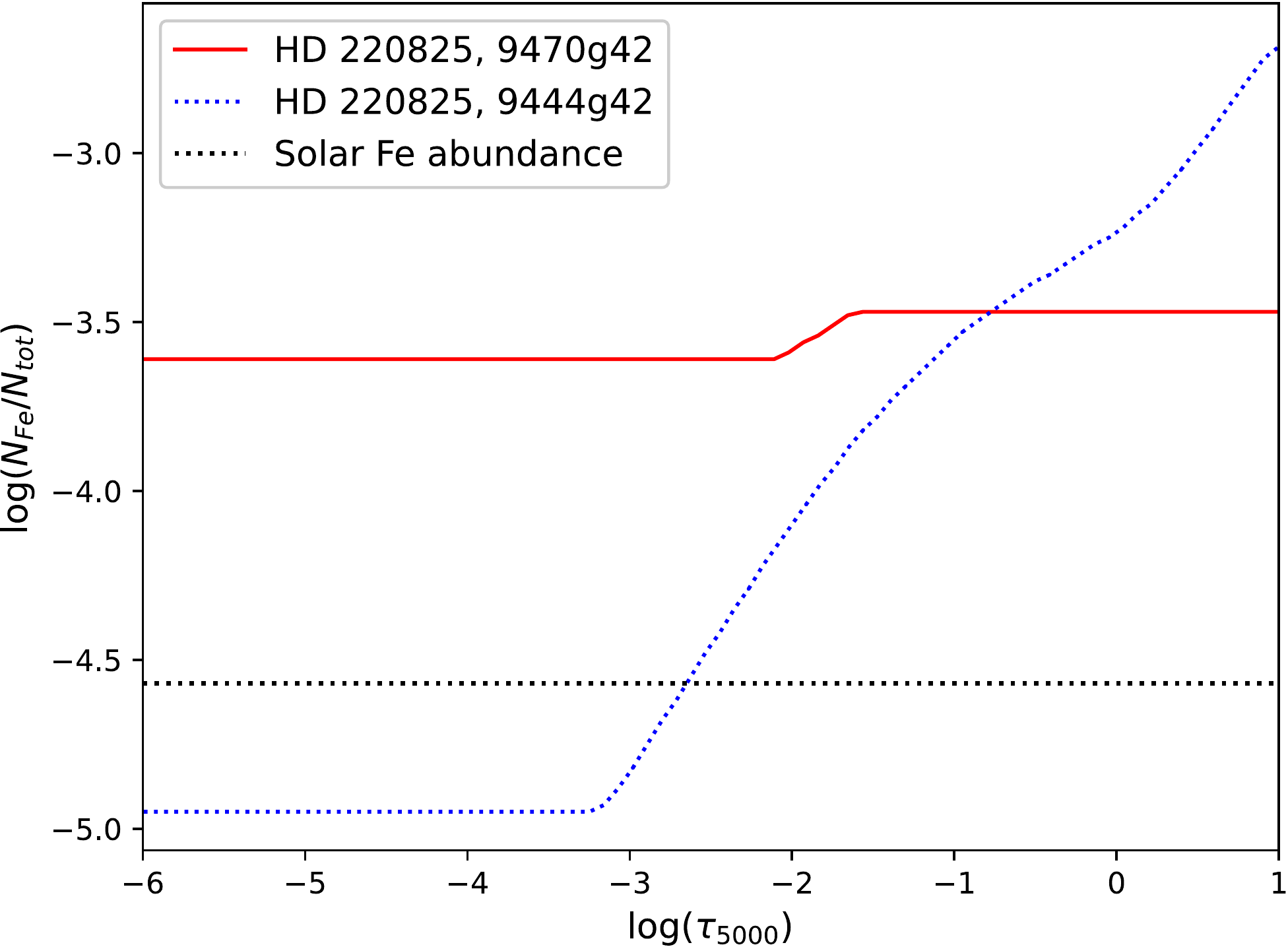}}\\a)
                \vfill
                \center{\includegraphics[width=1\linewidth]{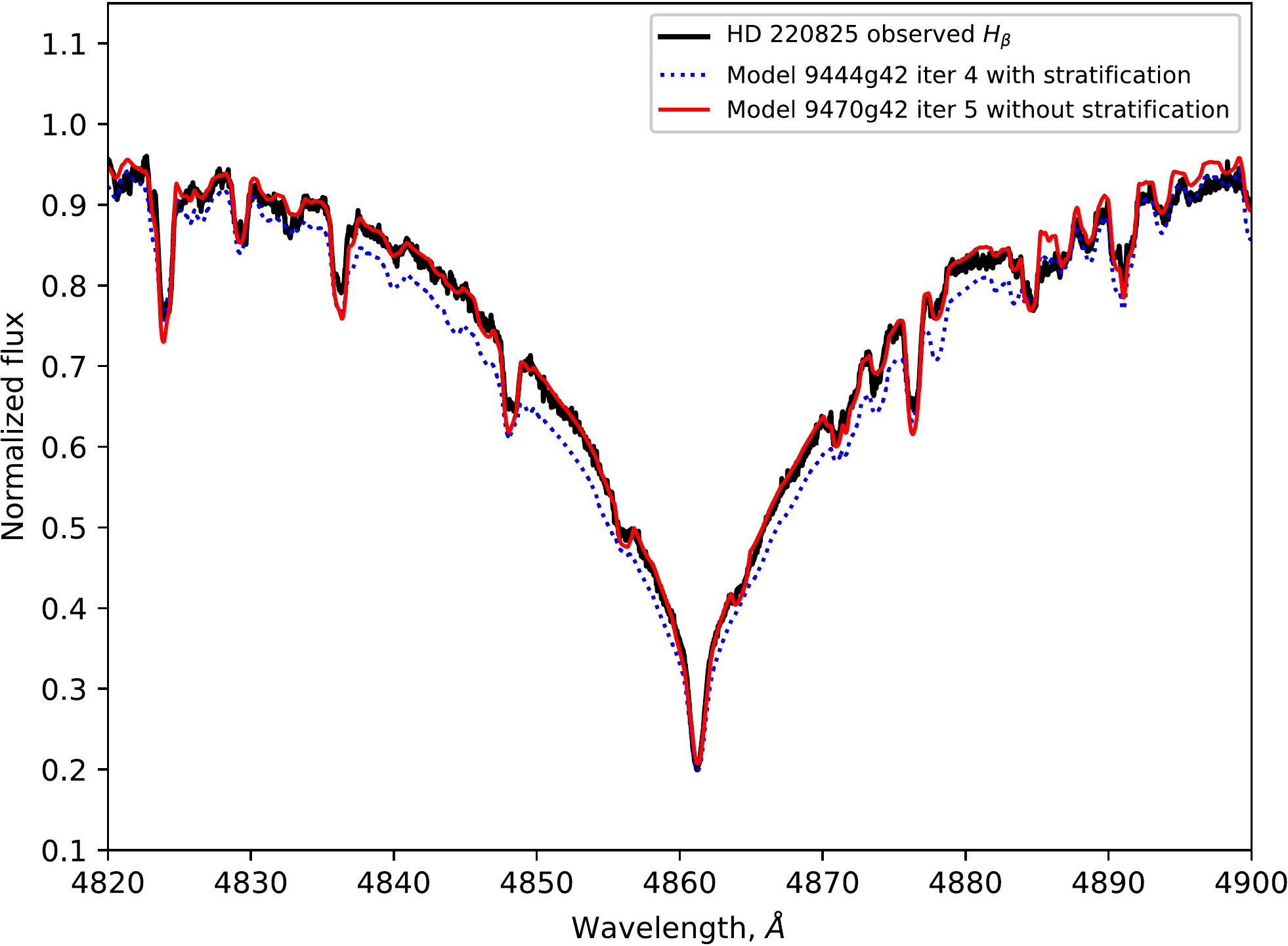}}\\b)
        \end{minipage}
        \hfill
        \begin{minipage}[h]{0.57\linewidth}
                \centering
                \includegraphics[width=1\linewidth, angle=90]{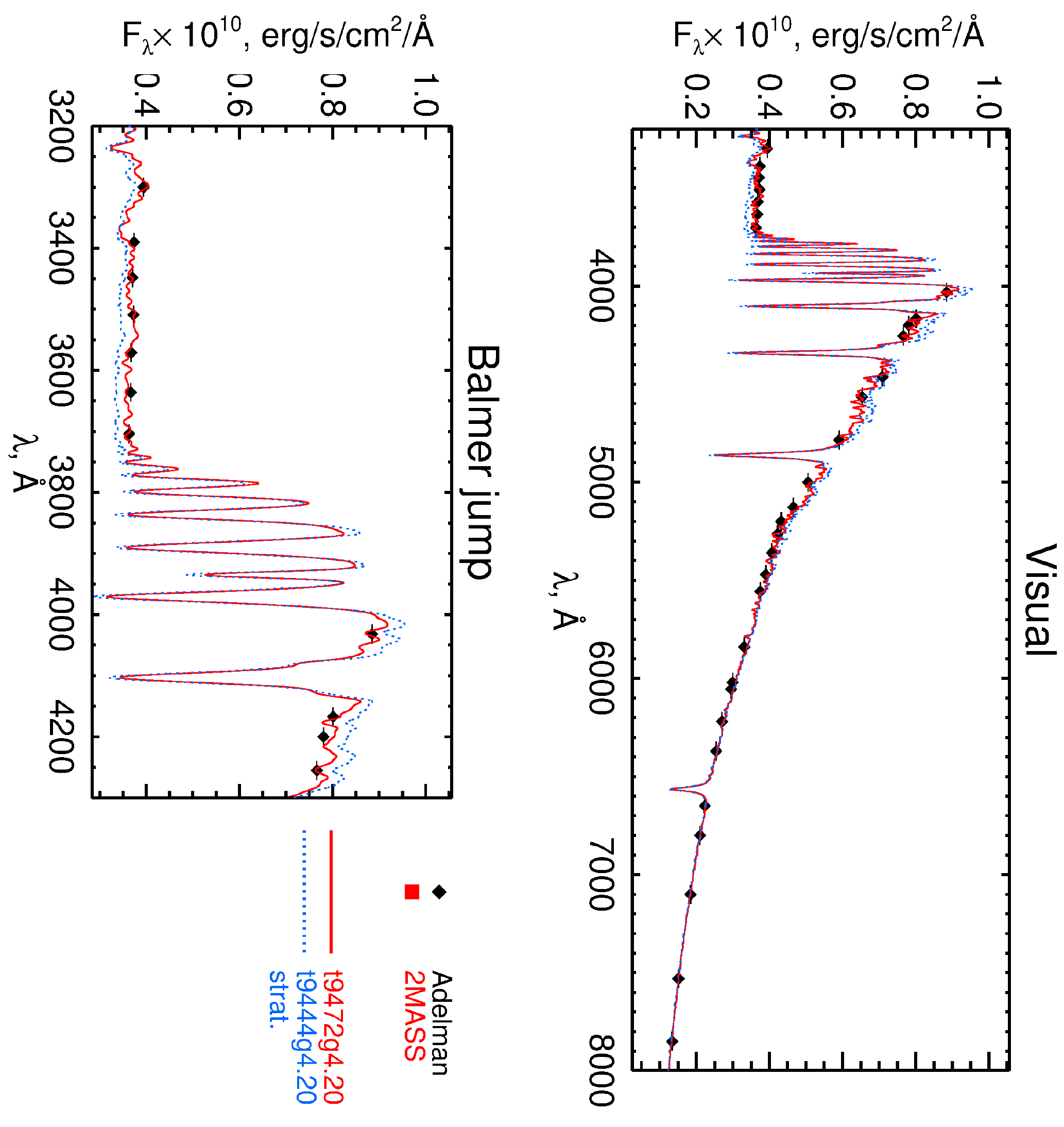}\\c)
        \end{minipage}
        \caption{a) Vertical abundance distribution (stratification) of Fe in the atmosphere of HD~220825 derived for model 9470g42 (red line) and for model 9444g42 (dotted blue line). The solar abundance is denoted by the dotted black line. b) Comparison of the observed $H_{\beta}$ line profile in the atmosphere of HD~220825 (black line) with theoretical profiles calculated assuming stratified (dotted blue line) and chemically homogeneous atmospheres (red line). c) Comparison between the observed and best-fit theoretical SEDs calculated with (dotted blue line) and without (solid red line) stratification. Observed SED data in the visual spectral region and around the Balmer jump are shown with filled black diamonds.}
        \label{HD220825-str}
\end{figure*}

A strong abundance gradient of these elements is observed mainly in cool stars (6000~K~--~8000~K), where the abundance jump amplitude decreases with increasing \teff\, and shifts to the upper atmospheric layers \citep{2009A&A...495..937L}. The stars in our sample have rather high effective temperatures and high rotational velocities \vsini, which causes the spectral lines to be broadened and blended. Because of this, it was difficult to perform an accurate stratification analysis.
Nevertheless, in order to estimate the possible presence of element stratification in our sample stars, we tried two sets of Fe lines (with a difference in two lines) for the analysis of Fe stratification in HD~220825. The resulting stratification profiles were derived using the code {\sc DDAFit} \citep{2005A&A...438..973R} and are shown in Fig.~\ref{HD220825-str}a. The stratification profiles are very different, even though the underlying models are almost similar: the difference in 26~K in their \teff\, is obviously too small to cause such a variation in the stratification profiles. The first set of lines results in a very much homogeneous Fe distribution (red curve), while using the second
set of Fe lines, we obtained a solution with large uncertainties for the abundance at low and upper atmospheric boundaries (blue curve). 
This second solution also resulted in a poorer fit to the hydrogen Balmer lines, as shown in Fig.~\ref{HD220825-str}b, which compares observed and predicted H$\beta$ line profiles
calculated assuming stratified and non-stratified models. Similar to the case of SED fitting (see next section), we found out that the chemically homogeneous model atmosphere describes the observed H$\beta$ line profile better than a stratified atmosphere. Therefore we finally chose models without stratification for the further analysis of our sample stars.

\section{Determining the fundamental parameters}\label{Model}

Our determination of fundamental parameters of stars is based on fitting the theoretical SED to the observed SED. We calculated the flux that emerges from the surface of a star using the code \textsc{LLmodels} . The distances to the stars are defined by stellar parallax values that we took from the GAIA DR2 catalogue \citep{2018yCat.1345....0G}. We also took interstellar absorption into account. HD~4778 is located at a distance of 107.268~pc from the Sun, and the correction for the interstellar reddening $A_v = 0.065$ ($E(B-V) = 0.02$) was applied \citep{2005AJ....130..659A}. The other stars, HD~120198 and HD~220825, are located closer to the Sun, which results in weaker interstellar absorption: $E(B-V) = 0.002(14)$ for HD~220825 and $E(B-V) = 0.008(11)$ for HD~120198 \citep{2014A&A...561A..91L}. Therefore we neglected the corrections for interstellar absorption for these two stars.

\begin{figure}[hbt!]
        \centering
        \includegraphics[width=0.9\linewidth, clip]{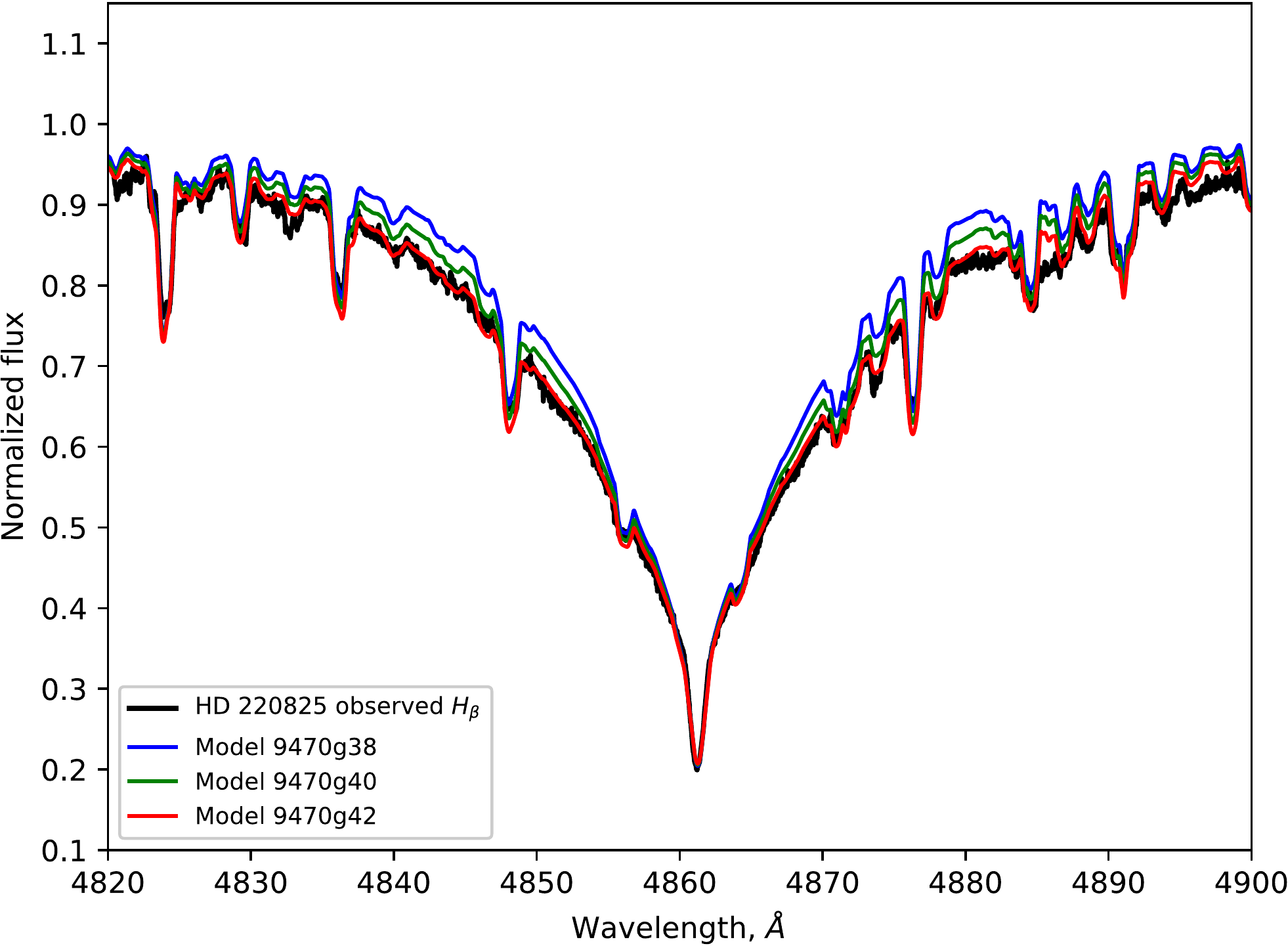}
        \caption{Comparison between the modelled and observed $H_{\beta}$ line profile in HD~220825 for models with same\teff\,= 9470 K and different \lgg\, = 3.8  (blue line), 4.0 (green line) and 4.2 (red line).}\label{Hbeta_HD220825}
\end{figure}

\begin{figure*}[hbt!]
        \centering
        \includegraphics[width=0.5\linewidth, angle=90, clip]{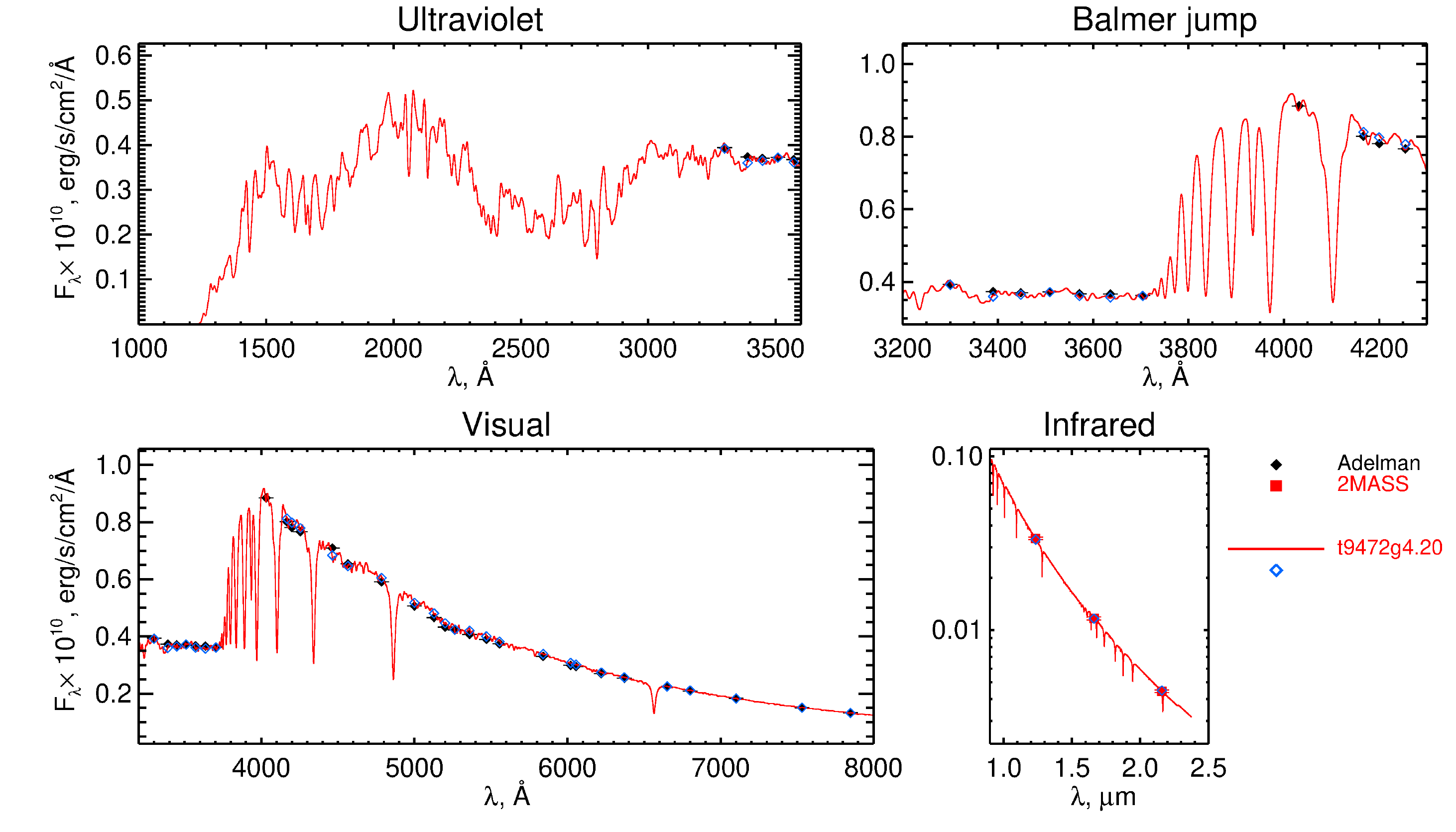}
        \caption{Comparison between the observed bution taken from the Adelman (filled black diamonds) and  2MASS (filled red squares) catalogues and best-fit theoretical flux (solid red line) calculated after five abundance analysis iterations for the \textsc{LLmodels} atmosphere of HD~220825 with the parameters \teff\, = 9470 K and \lgg\, = 4.2. Open blue diamonds show the theoretical fluxes convolved with the corresponding filters that were used in actual observations.}\label{SED-hd220825}
\end{figure*}

\begin{figure*}[hbt!]
        \centering
        \includegraphics[width=0.5\textwidth, angle=90, clip]{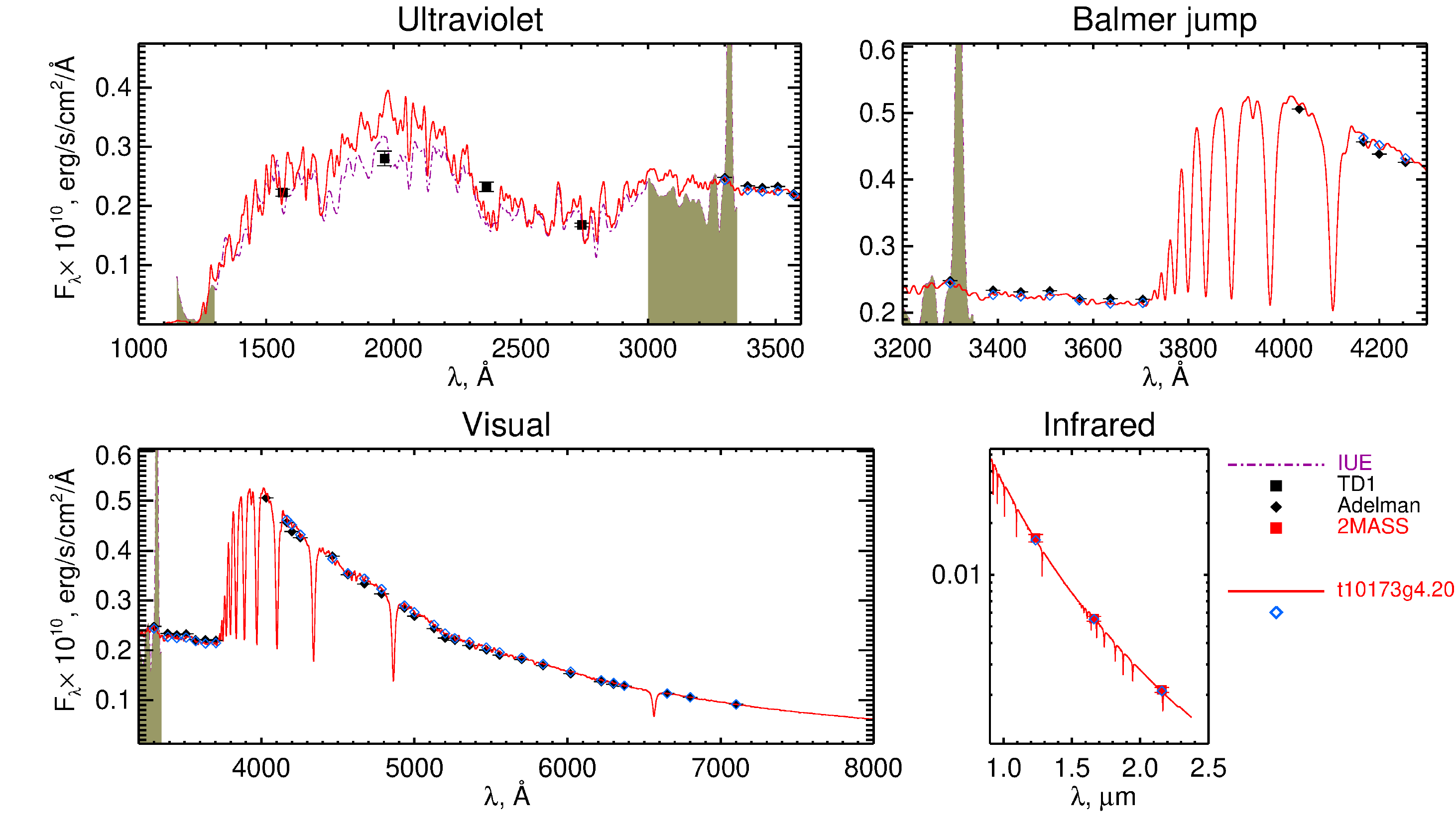}
        \caption{Comparison of the observed energy distribution from Adelman (filled black diamonds), TD1 (filled black squares) and 2MASS photometric data (filled red squares) with the best-fit theoretical flux (solid red line) calculated after five abundance analysis iterations for the \textsc{LLmodels} atmosphere of HD~120198 with the parameters \teff\, = 10173 K and \lgg\, = 4.2. Open blue diamonds show the theoretical fluxes convolved with the corresponding filters. The shaded area illustrates the spectral regions that were excluded from the fit.}\label{HD120198-sed}
\end{figure*}

\begin{figure*}[hbt!]
        \centering
        \includegraphics[width=0.5\linewidth, angle=90, clip]{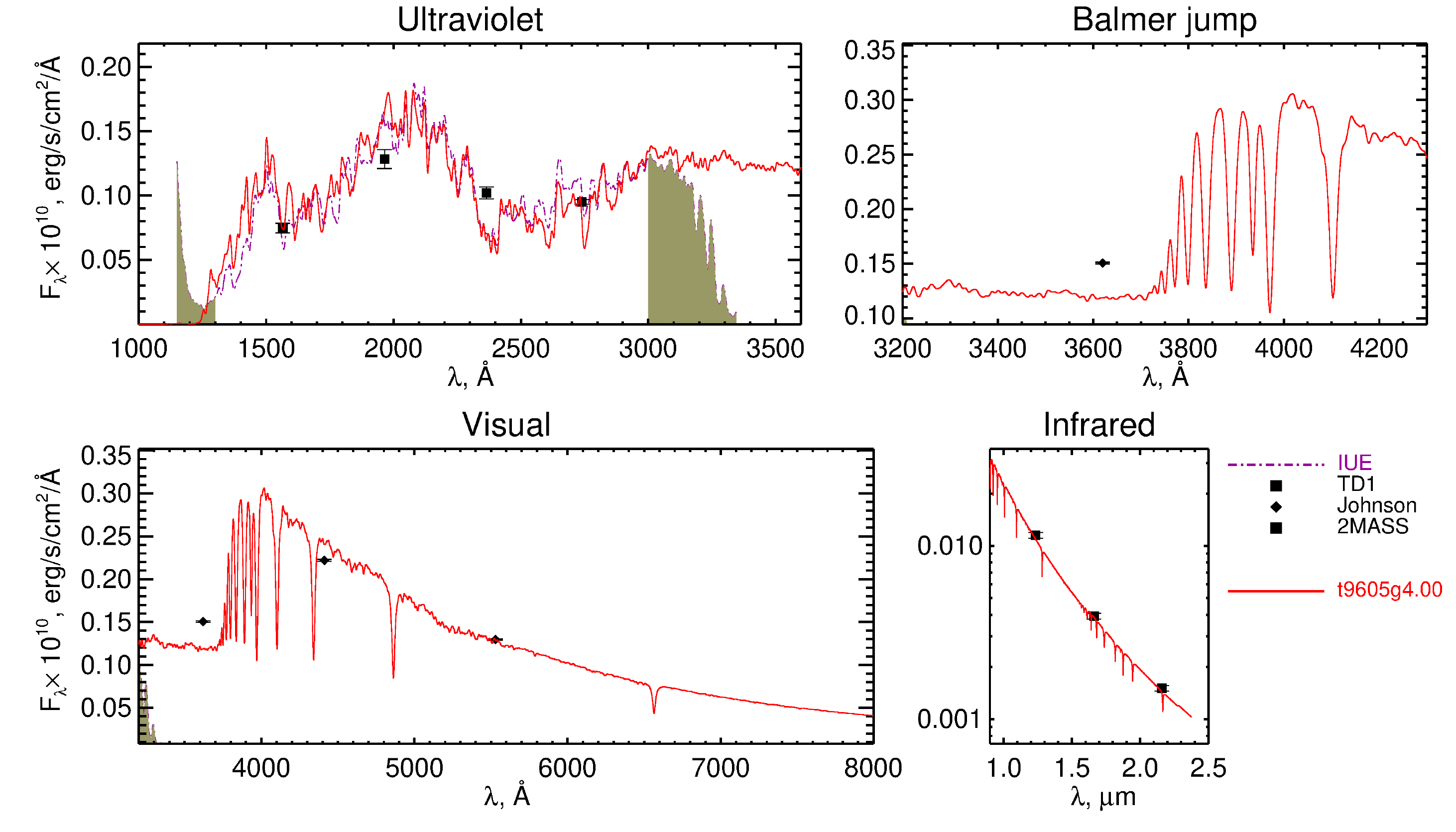}
        \caption{Comparison of the observed energy distribution from Johnson (filled black diamonds), TD1 (filled black squares), and 2MASS photometry (filled black squares in Infrared) with the best-fit theoretical flux (solid red line) calculated after three abundance analysis iterations for the \textsc{LLmodels} atmosphere of HD~4778 with the parameters \teff\, = 9605 K and \lgg\, = 4.0. The shaded area illustrates spectral regions that were excluded from the fit.}\label{SED-hd4778-elodie}
\end{figure*}

We compared the theoretical flux calculated with our model atmosphere code with the observed SED. For a given abundance pattern, we optimised \teff, \lgg, and the stellar radii R to obtain their best-fit values. The final fits are shown in Fig.~\ref{SED-hd220825} (HD~220825), Fig.~\ref{HD120198-sed} (HD~120198), and Fig.~\ref{SED-hd4778-elodie} (HD~4778). We found out that in all cases, the values of \lgg\, could not be robustly determined from the fit to the observed SED. Therefore we determined \lgg\, by  fitting the hydrogen line profiles (see e.g. Fig.~\ref{Hbeta_HD220825}). For HD~4778, the value of \lgg=4.0 was taken, which resulted in the final best-fit model of 9605g40. For this star, we were able to construct a model atmosphere even without spectrophotometric data in the optical region, using Johnson's photometry alone.

Our final uncertainties on the stellar radii account for errors in both \teff\, and parallax (distance) for all stars. The final fundamental parameters of the program stars are collected in Table~\ref{Fund_param_ApStars}.

\subsection{Effect of surface abundance inhomogeneity on the determination of fundamental parameters}

 The determination of fundamental parameters presented in our paper is based on fitting to the time-averaged spectrophotometric scan from Adelman's catalogue.
 However, Ap stars are known to be photometrically and spectroscopically variable due to the presence of abundance spots on their surfaces. These spots lead to the modulation
 of the SED as the star rotates and potentially cause biases in the derived parameters such as \teff. 
 To study the effects of surface inhomogeneity, we investigated the case of HD~220825, a star that clearly shows a spotted surface \citep{1996AstL...22..822R} that can lead to light variability \citep[see e.g.][]{2010A&A...524A..66S}.

  Because we have spectrophotometric observations by Adelman that were obtained at rotation phases different from the ESPaDOnS spectrum that we used for the abundance analysis (rotation phase 0.011), it allowed us to estimate possible errors in \teff\, and \lgg\, associated with an inhomogeneous surface abundance distribution. Spectriphotometric scans in ten rotation phases were extracted from  \citet{1983A&AS...51..365P}. Table~\ref{TeffAdelman} provides \teff\, values and their errors  derived by fitting to these individual spectrophotometric scans. 
 \lgg\, was difficult to estimate unambiguously, and we would simply set a rather conservative uncertainty of 0.2 dex (\lgg\, = 3.8, 4.0, and 4.2 were used to choose a model atmosphere by fitting hydrogen line profiles).

\begin{table}[hbt!]
        \label{TeffAdelman}
        \caption{\teff\, derived from fitting spectrophotometric scans obtained in ten rotation phases.}
        \footnotesize
        \centering
        \begin{tabular}{l c c c }
                \hline HJD & Phase & \teff, K & Error \\
                \hline
                2441559.0 & 0.895 & 9415 & 94 \\
                2441561.0 & 0.385 & 9429 & 93 \\
                2441561.8 & 0.934 & 9545 & 37 \\
                2441563.0 & 0.703 & 9507 & 37 \\ 
                2441577.8 & 0.234 & 9488 & 34 \\
                2441578.8 & 0.942 & 9522 & 31 \\ 
                2441616.8 & 0.717 & 9451 & 25 \\
                2443061.8 & 0.587 & 9477 & 26 \\ 
                2443063.8 & 0.941 & 9498 & 23 \\
                2443066.8 & 0.985 & 9431 & 22 \\
                \hline&  Average & 9476 & 41 \\
                \hline
        \end{tabular}
        \tablefoot{The averaged \teff\, with its standard deviation is given in the last row.}
\end{table}

\begin{figure}[hbt!]
        \centering
        \includegraphics[width=0.6\linewidth, angle=-90, clip]{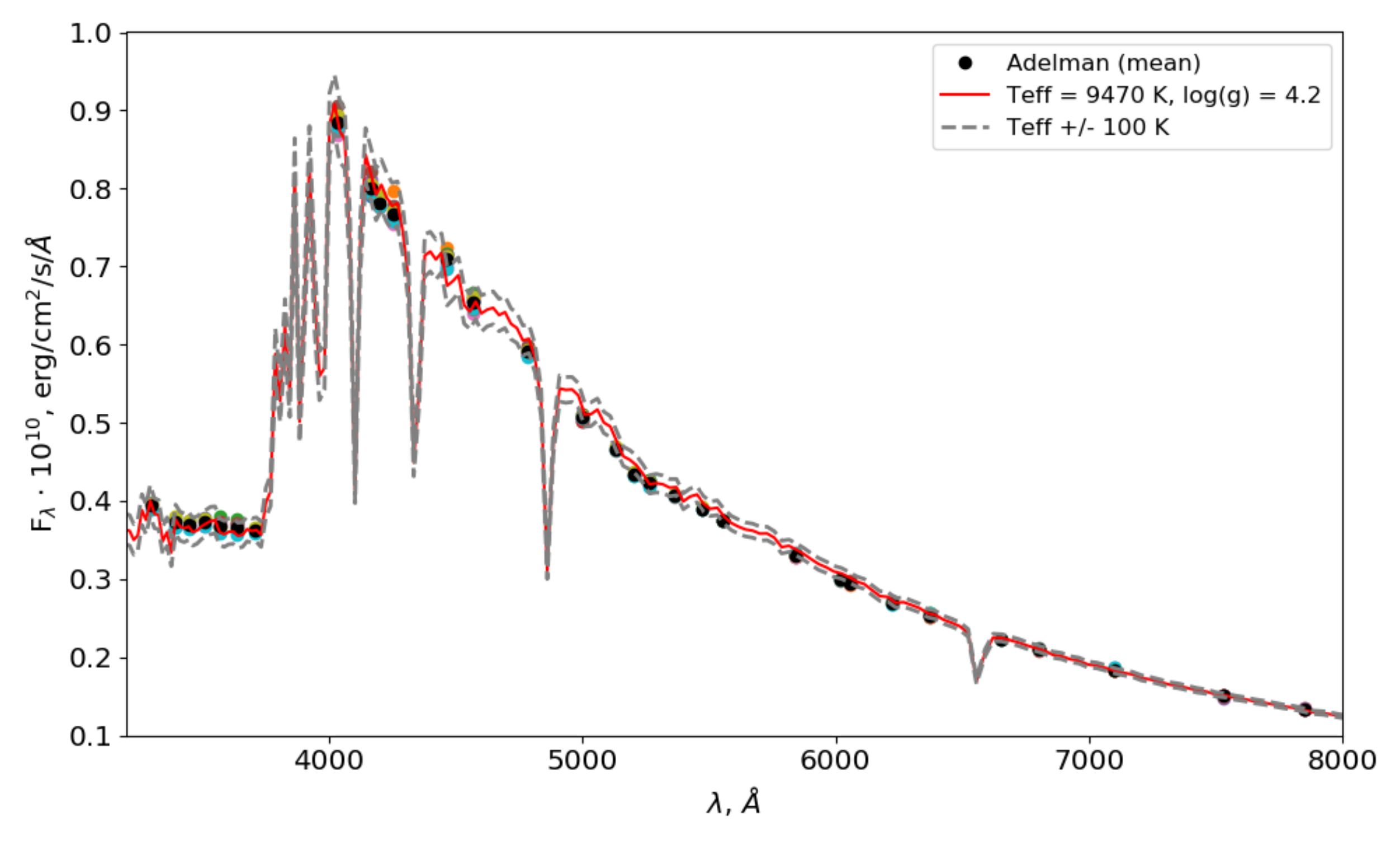}\\
        \caption{Comparison between the observed Adelman spectrophotometric scans (colour symbols) and theoretical SED for HD~220825. The solid red line shows the theoretical energy distribution computed for the final self-consistent model atmosphere of HD~220825 with parameters \teff\, = 9470 K and \lgg\, = 4.2. The dashed grey lines illustrate the effect of changing \teff\, by $\pm$100 K. The averaged spectrophotometric scan is represented by filled black circles.}\label{SED_HD220825_adelman}
\end{figure}

To estimate the effect of the flux variability caused by surface abundance spots, we plot an averaged spectrophotometric scan (filled black circles) together with the scans obtained at different phases (filled coloured circles) in Fig.~\ref{SED_HD220825_adelman}. We compared observations and the theoretical SED calculated with the finally adopted parameters for HD~220825: \teff\, = 9470 and \lgg\, = 4.2. In order to see the effect of variability on the determination of stellar parameters, we calculated the models with the parameters \teff\, = $9470\pm100$~K and \lgg\, = 3.8, 4.0. The temperature variations within $\pm100$~K are enough to fit all observations. At the same time, changing \lgg\, has only little effect on the SED. We thus conclude that the flux variations caused by the surface inhomogeneity do not lead to an effective temperature change by more than $\pm100$~K, which is a typical error in temperature estimates of Ap stars in similar self-consistent studies. 

\section{Comparison with the interferometry}\label{Comp}
The current study completes a programme of fundamental parameter determinations by means of spectroscopy of a  sample of Ap stars that were chosen for interferometric measurements. The final sample of stars with fundamental parameters obtained from interferometric observations is given in \citet{2020A&A...642A.101P}. It contains 14 stars, and for all but one of them, the self-consistent spectroscopic modelling was carried out. First, we compare the spectroscopic and interferometric results obtained in this work.

\begin{figure*}[hbt!]
\begin{minipage}{0.33\textwidth}
  \centering
  \includegraphics[width=1.0\textwidth,clip]{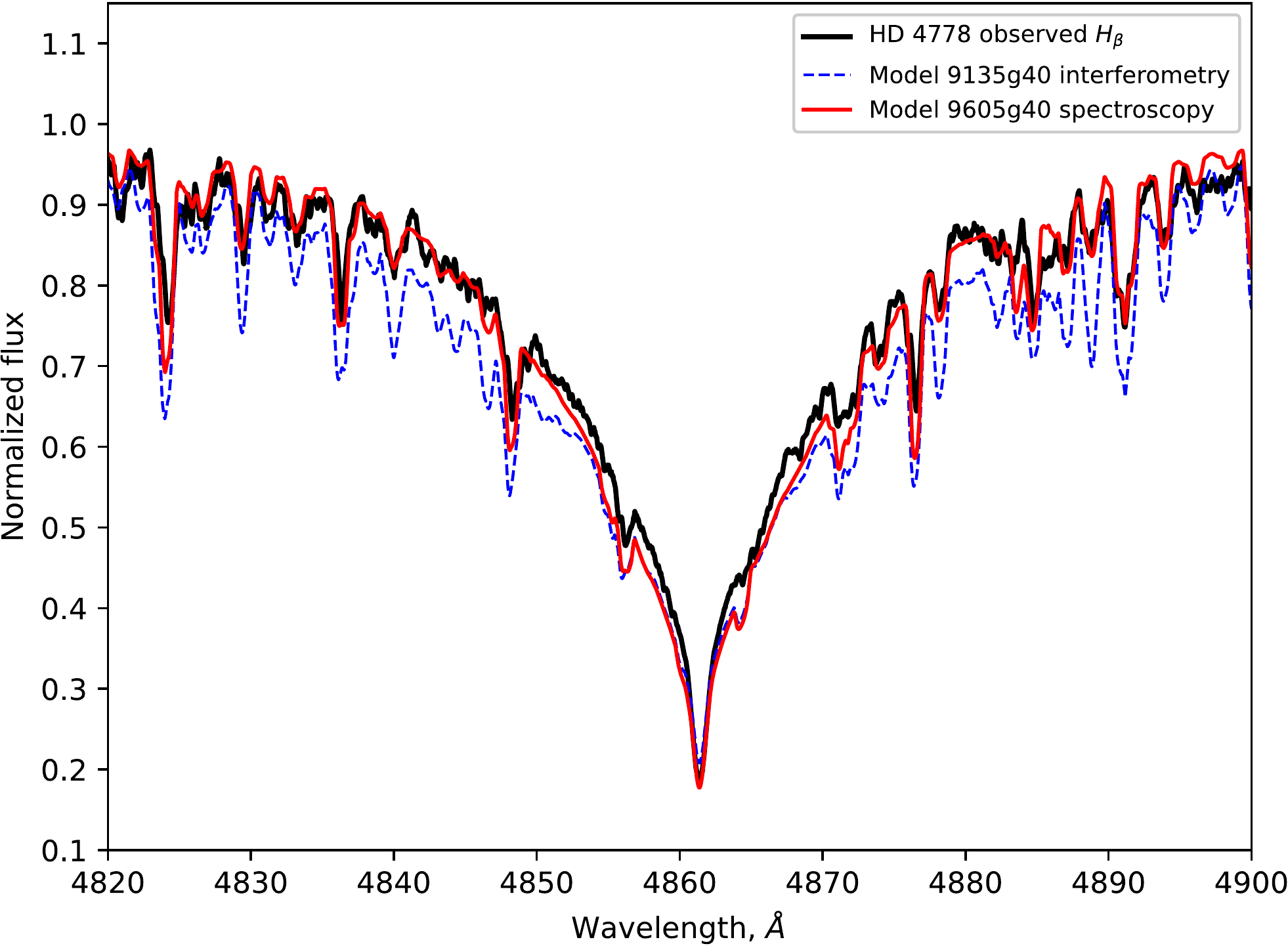}\\a)
\end{minipage}
\begin{minipage}{0.33\textwidth}
  \centering
  \includegraphics[width=1.0\textwidth,clip]{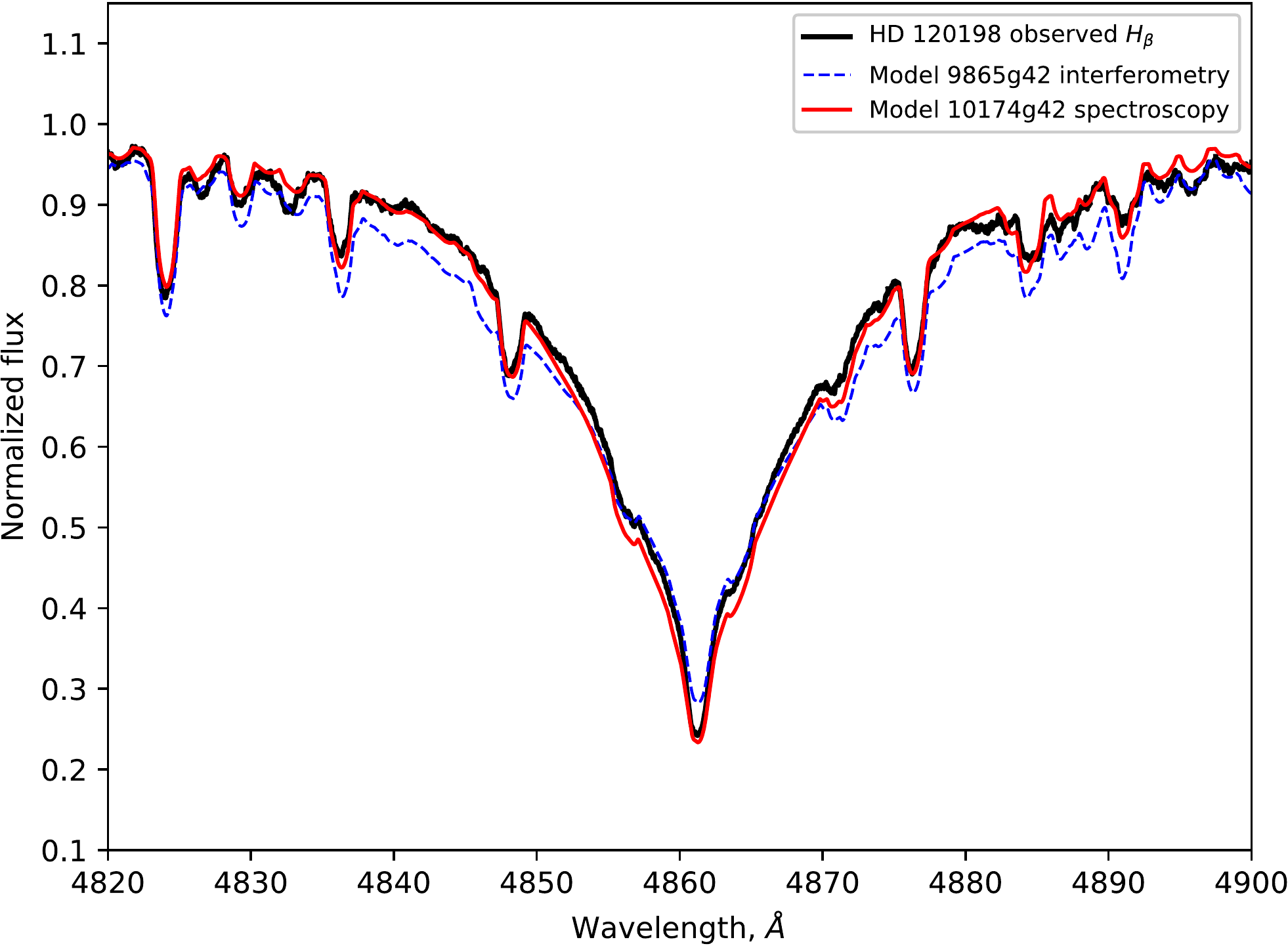}\\b)
\end{minipage}
\begin{minipage}{0.33\textwidth}
  \centering
  \includegraphics[width=1.0\textwidth,clip]{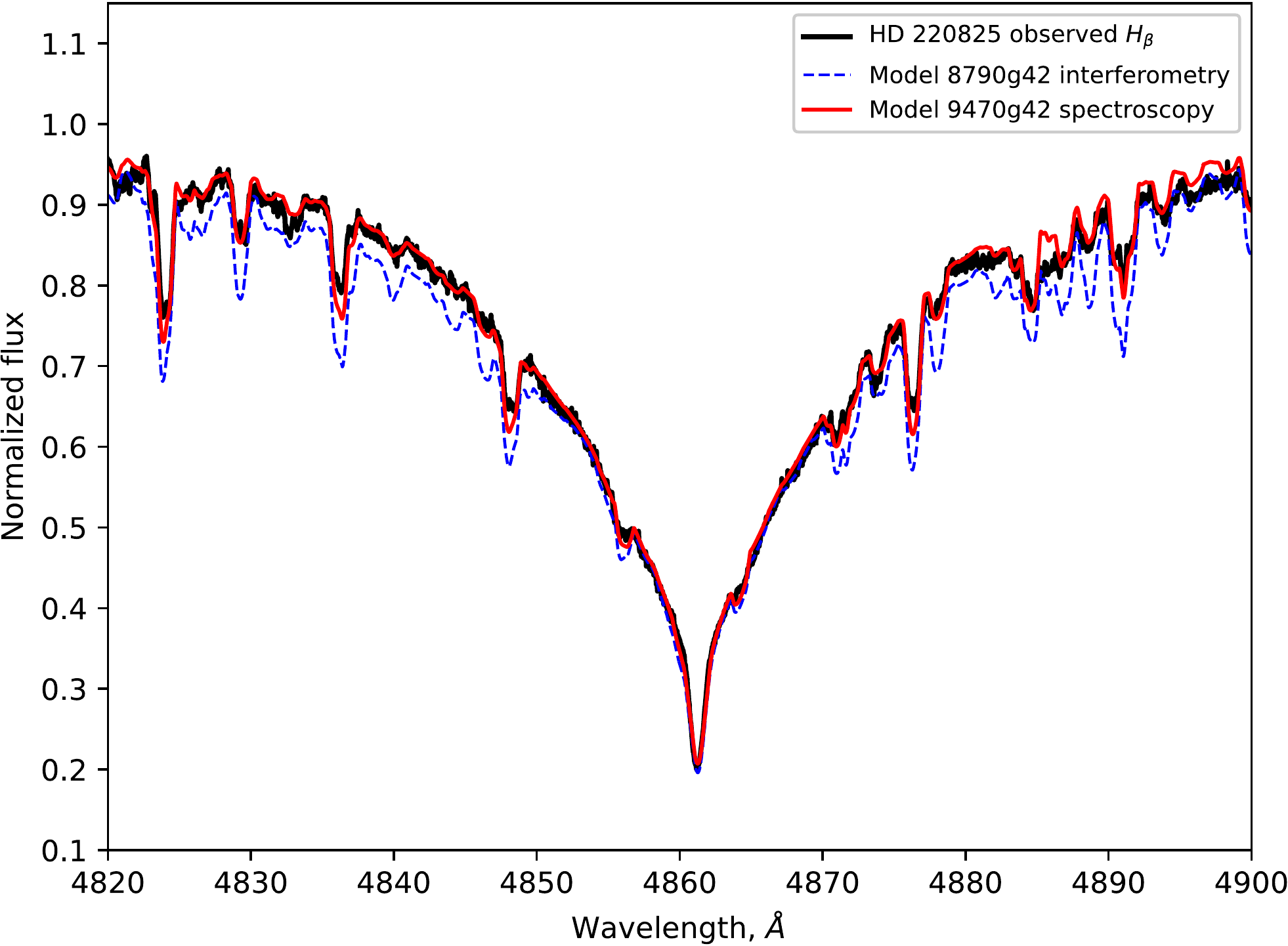}\\c)
\end{minipage}
\caption{Comparison between the modelled and observed $H_{\beta}$ line profile in (a) HD~4778, (b) HD~120198, and (c) HD~220825 for models with the same \lgg\, and different \teff\, obtained by various methods: \teff\, = 9605 K (HD~4778), 10174 K (HD~120198), 9470 K (HD~220825) by self-consistent modelling (red line) and \teff\, = 9135 K (HD~4778), 9865 K (HD~120198), 8790 K (HD~220825) by interferometry (dashed blue line).}\label{stars-hbeta}
\end{figure*}

\underline{HD~4778.}  
The obtained \teff$^{interf}$ is lower than \teff$^{spec}$. However, the latter agrees well with the photometric analysis within the errors. We thus tried to fix the temperature using an additional diagnostic, which was to fit the $H_{\beta}$ line profile with model atmospheres calculated assuming both \teff\ estimates. The comparison is presented in Fig.~\ref{stars-hbeta}a. For both temperatures, the synthetic spectra were calculated with the same abundances. The model atmosphere with the spectroscopic temperature fits the $H_{\beta}$ line profile noticeably better than the interferometric temperature. 

\underline{HD~120198.}
While the fundamental parameters of HD~120198 agree within the errors of the determination, \teff\, estimated by spectroscopy is higher  and fits the observed H$\beta$ line better than  the interferometric \teff\,  (Fig.~\ref{stars-hbeta}b).

\begin{figure*}[hbt!]
    \begin{minipage}{0.33\textwidth}
        \centering
            \includegraphics[width=1.0\textwidth,clip]{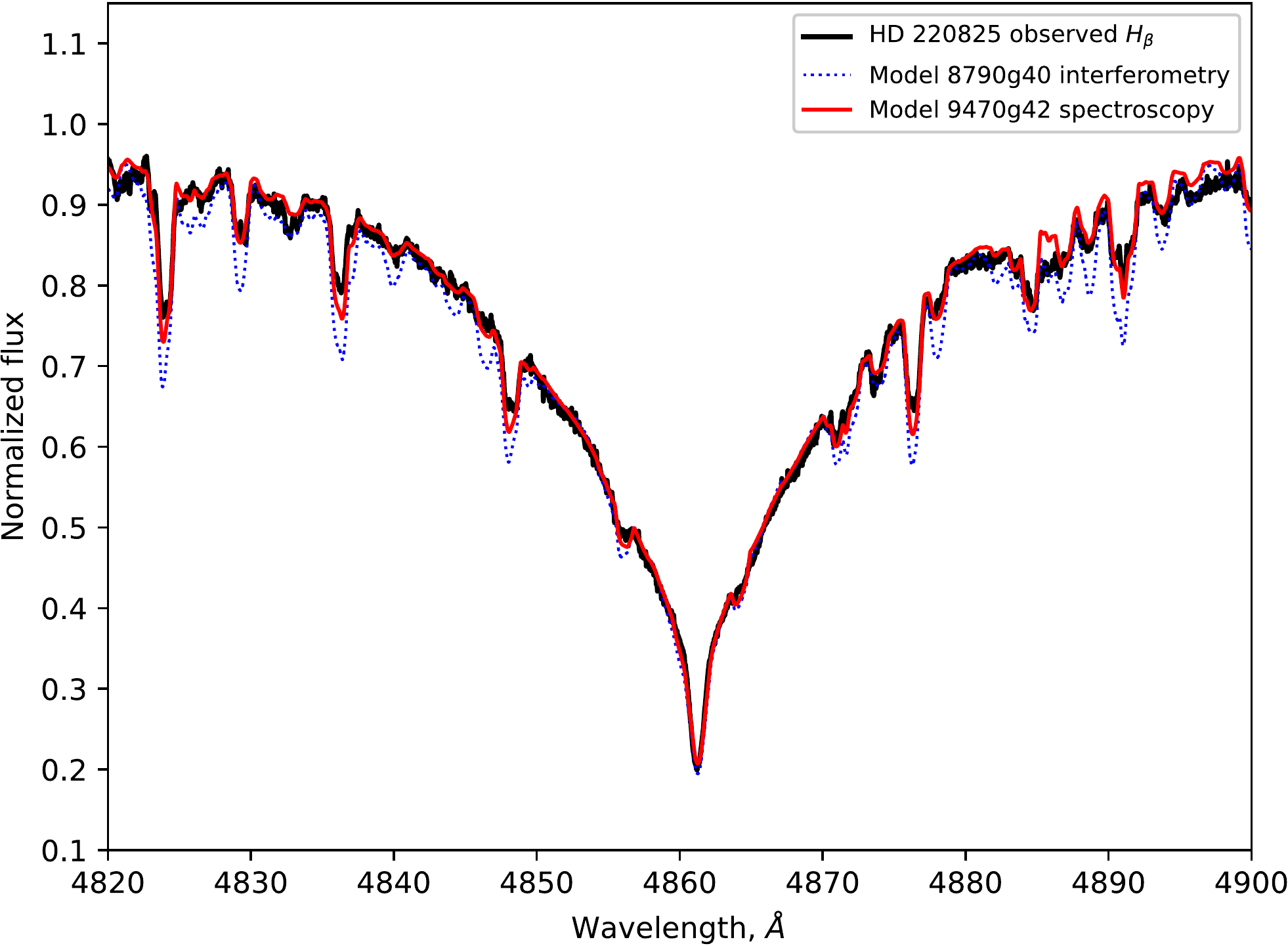}\\a)
    \end{minipage}
    \begin{minipage}{0.65\textwidth}
        \centering
            \includegraphics[width=0.38\textwidth, angle=90, clip]{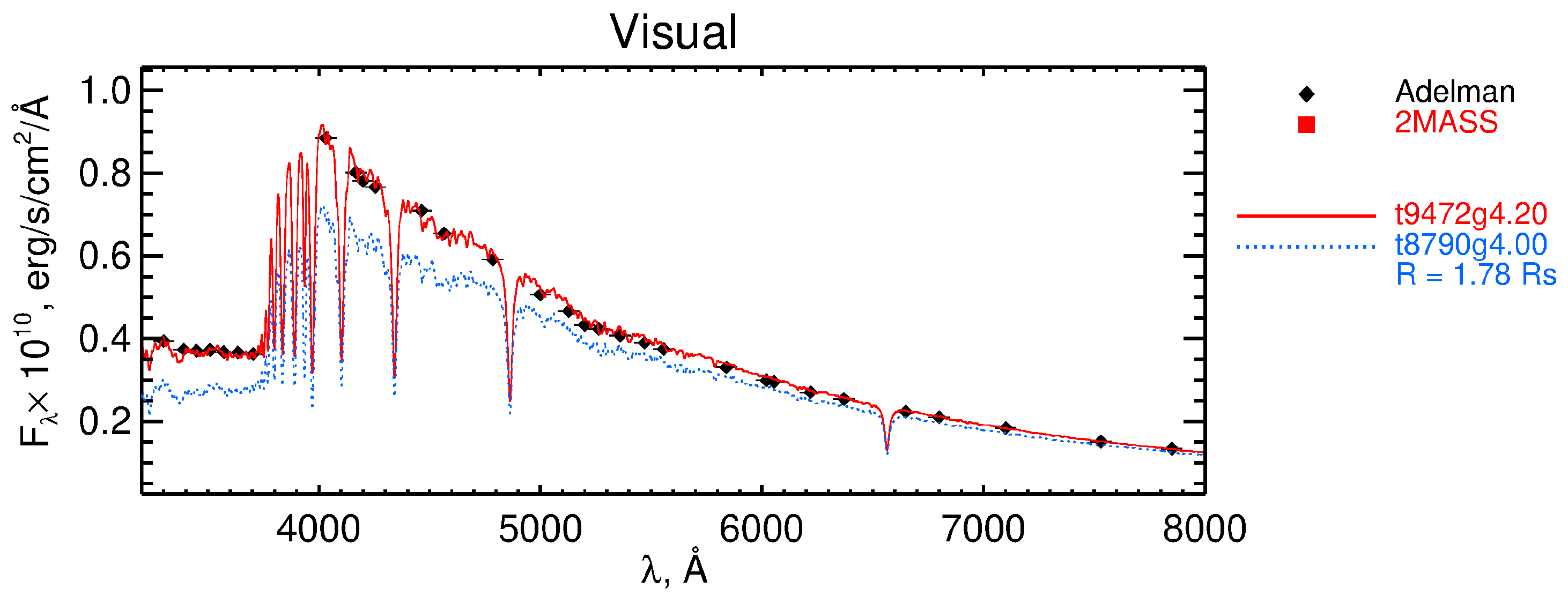}\\b)
    \end{minipage}
        \caption{Comparison between the observed $H_{\beta}$ line profile (a) and Adelman's spectrophotometric scan (b) and model fits calculated with spectroscopic (red lines) and interferometric \teff\ of HD~220825 (dotted blue lines).}\label{HD220825_comp}
\end{figure*}

\underline{HD~220825.} 
For this star, we obtained the largest difference between the effective temperatures derived through spectroscopy and interferometry. The difference exceeds the quoted errors. The radii of this star obtained with two methods are very close ($R/R_{\odot}^{spec} = 1.71$ and $R/R_{\odot}^{interf} = 1.78$), therefore the lower effective temperature derived based on the interferometric analysis might be a result of the total flux underestimation by \citet{2020A&A...642A.101P}. There are no flux measurements of HD~220825 in the UV region, which forced \citet{2020A&A...642A.101P} to estimate this flux by approximate correlations between flux in optical and UV deduced from the observations of other stars. We note that the spectroscopic temperature \teff=9470~K fits the H$\beta$ line profile much better than the interferometric temperature \teff=8790~K, assuming the same \lgg=4.2 for both cases (see Fig.~\ref{stars-hbeta}c). However, when the value of the surface gravity is decreased from \lgg=4.2 to \lgg=4.0, then we obtain a reasonable fit to the observed H$\beta$ line profile with the interferometric temperature ( Fig.~\ref{HD220825_comp}a). A fit to the SED does not provide a robust estimate for \lgg\ in the temperature range 9000~K--10000~K (see Section~\ref{Model}), and a decrease in \lgg\ does not noticeably change the spectrophotometric fit, but a combination of the interferometric temperature with any of the \lgg-values does not fit the observed spectrophotometry, as demonstrated in  Fig.~\ref{HD220825_comp}b.

\begin{figure}[hbt!]
\centering
         \includegraphics[width=0.45\linewidth, angle=90, clip]{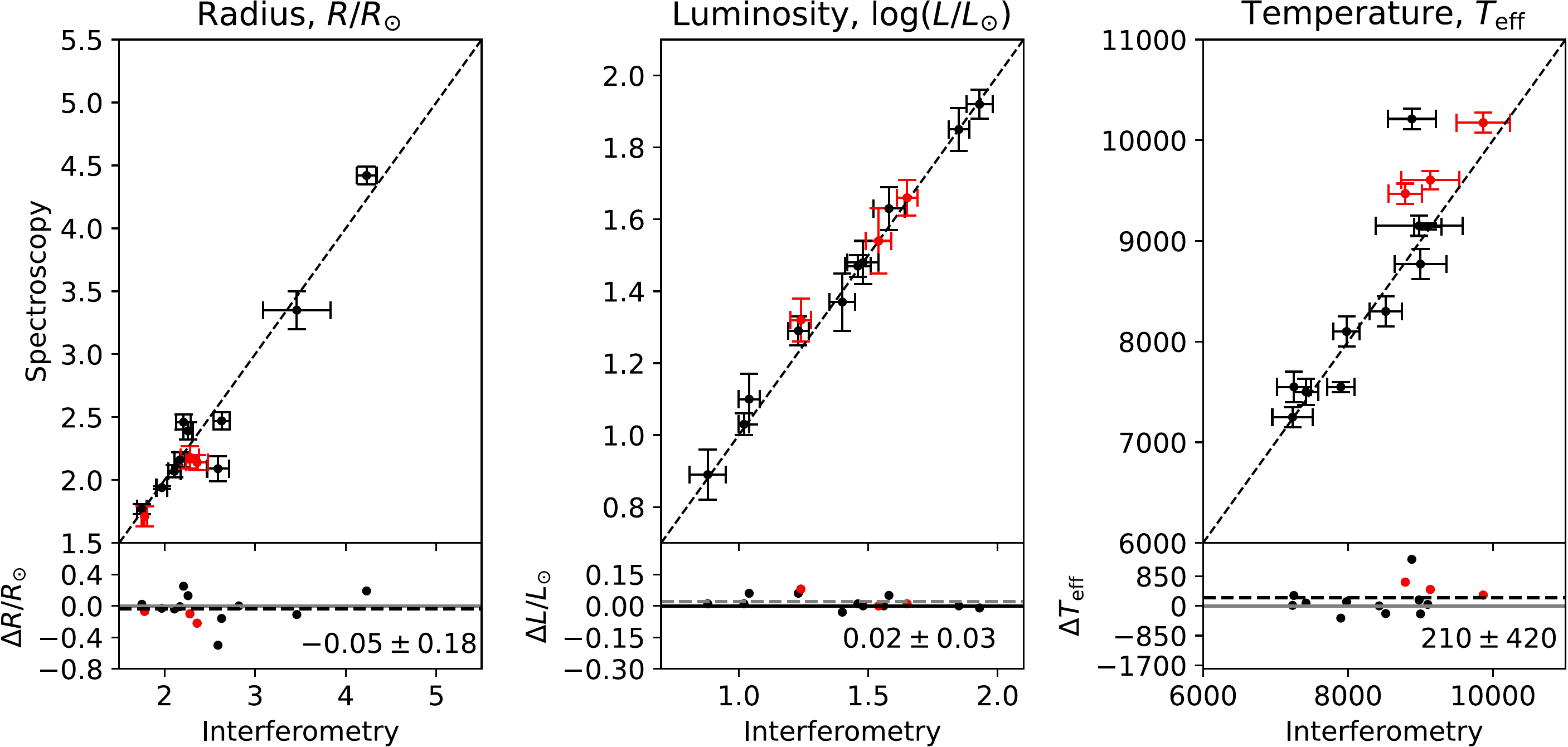}\\
         \includegraphics[width=0.45\linewidth, angle=90, clip]{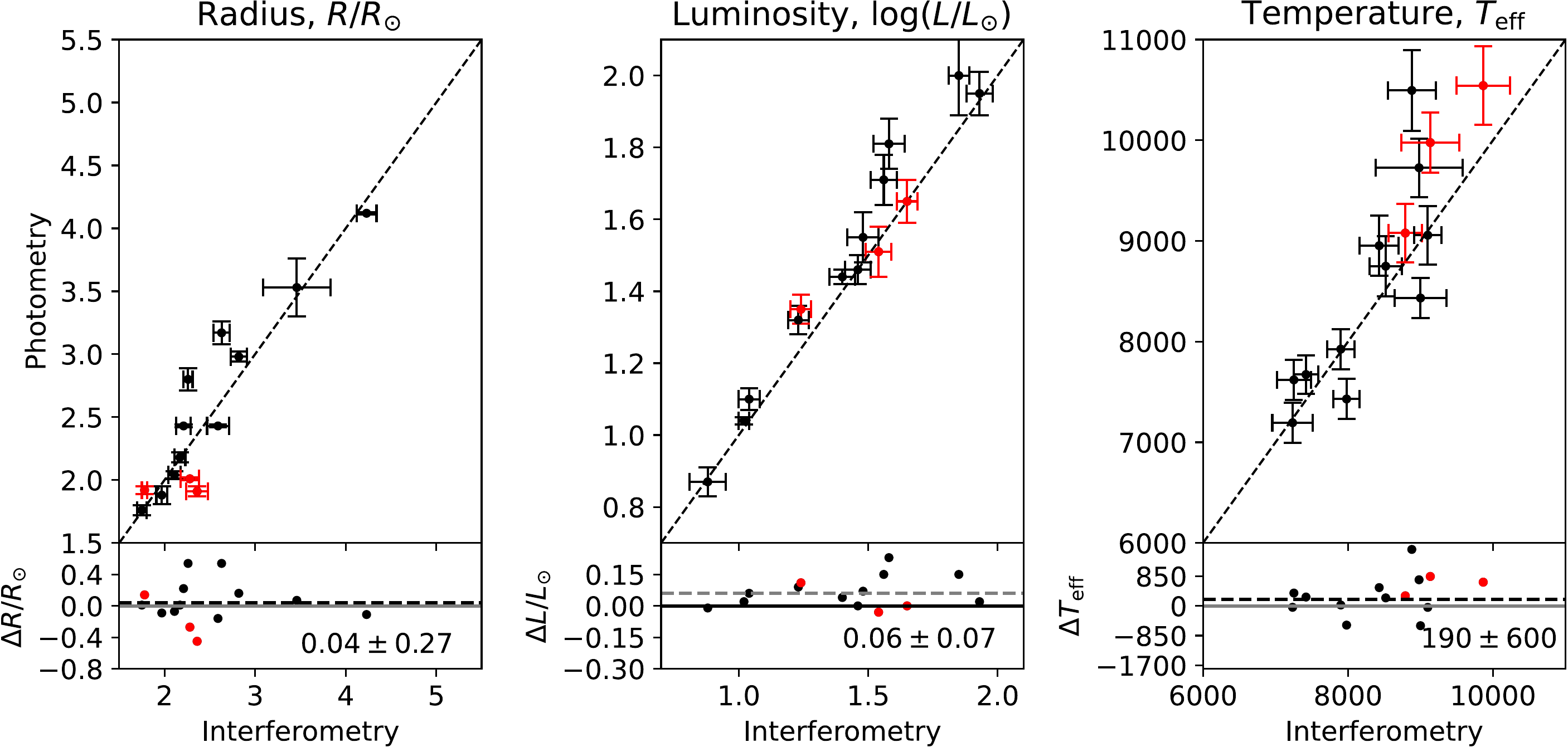}
        \caption{Comparison between radii, luminosity, and effective temperatures obtained by spectroscopy,  photometry, and interferometry. Red dots represent the stars under study.}\label{sp-phot-interf}
\end{figure}

Table~\ref{Fund_param_ApStars} collects the fundamental parameters for 13 benchmark stars obtained by interferometric \citep{2020A&A...642A.101P}, photometric \citep{2006A&A...450..763K}, and spectroscopic methods. The parameters derived by indirect methods (spectroscopy and photometry) are plotted versus those derived by the direct method (interferometry) in Fig.~\ref{sp-phot-interf}. The comparison of the parameters obtained by spectroscopy and interferometry shows a rather good agreement between them for all stars: the spectroscopic radii are smaller by 5\%\ on average than the interferometric radii, which is within 2$\sigma$ of the interferometric measurements. The luminosity agrees even better. The fundamental parameters derived from photometric calibrations do not show significant systematic shifts relative to spectroscopic and thus interferometric parameters in general. However, the dispersion in the parameter correlations is approximately twice larger than in spectroscopy. Our comparison shows that for stars with \teff$>$9000~K, the spectroscopically
derived temperatures are often significantly higher than the interferometric estimates. This is the case for HD~220285 and HD~4778, and also for HD~108662 
(see Table~\ref{Fund_param_ApStars}).
The reason might be inaccurate radius determinations by interferometry. For instance, we note that the difference between spectroscopic and interferometric determinations is higher for stars with smaller angular diameters $\theta<0.3$~mas on average, which is very close to the current instrument limit.
The largest difference is found for HD~108662, where the overestimation of stellar radii by interferometric measurements results in a \teff\, that is more than 1300~K lower
than the spectroscopic estimate. In the case of HD~220285, alternatively, the lack of observed flux in the UV region in which A-type stars radiate most of their energy might bias interferometric estimates towards lower \teff\, values. Conversely, a model fit to the observed SED does not strongly rely on the availability of observations in all spectral regions. It is mandatory, however, that all available flux observations are consistently fitted with a single model atmosphere, as we did in this study for HD~4778, where in the visual domain only two photometric flux measurements were available. The resulting \teff\, is then set by the model input and not by wavelength integration of the observed flux. Using model atmospheres also allows us to detect a possible inconsistency between the flux observed in different wavelength domains and instruments or missions, as was the case for UV observations of HD~108662, for instance \citep{2020INASR...5..219R}. The independent \teff\, diagnostic (hydrogen Balmer lines) favours spectroscopically derived values for all these cases.

This comparison shows that spectroscopic methods can be successfully used to measure stellar radii with sufficient accuracy for the Ap stars for which interferometric observations are still impossible. We also showed that a less time-consuming photometric method provides stellar parameters that are accurate enough to be used for statistical studies of a large sample of Ap stars.

\begin{figure}[hbt!]
\centering
         \includegraphics[width=0.7\linewidth, angle=90, clip]{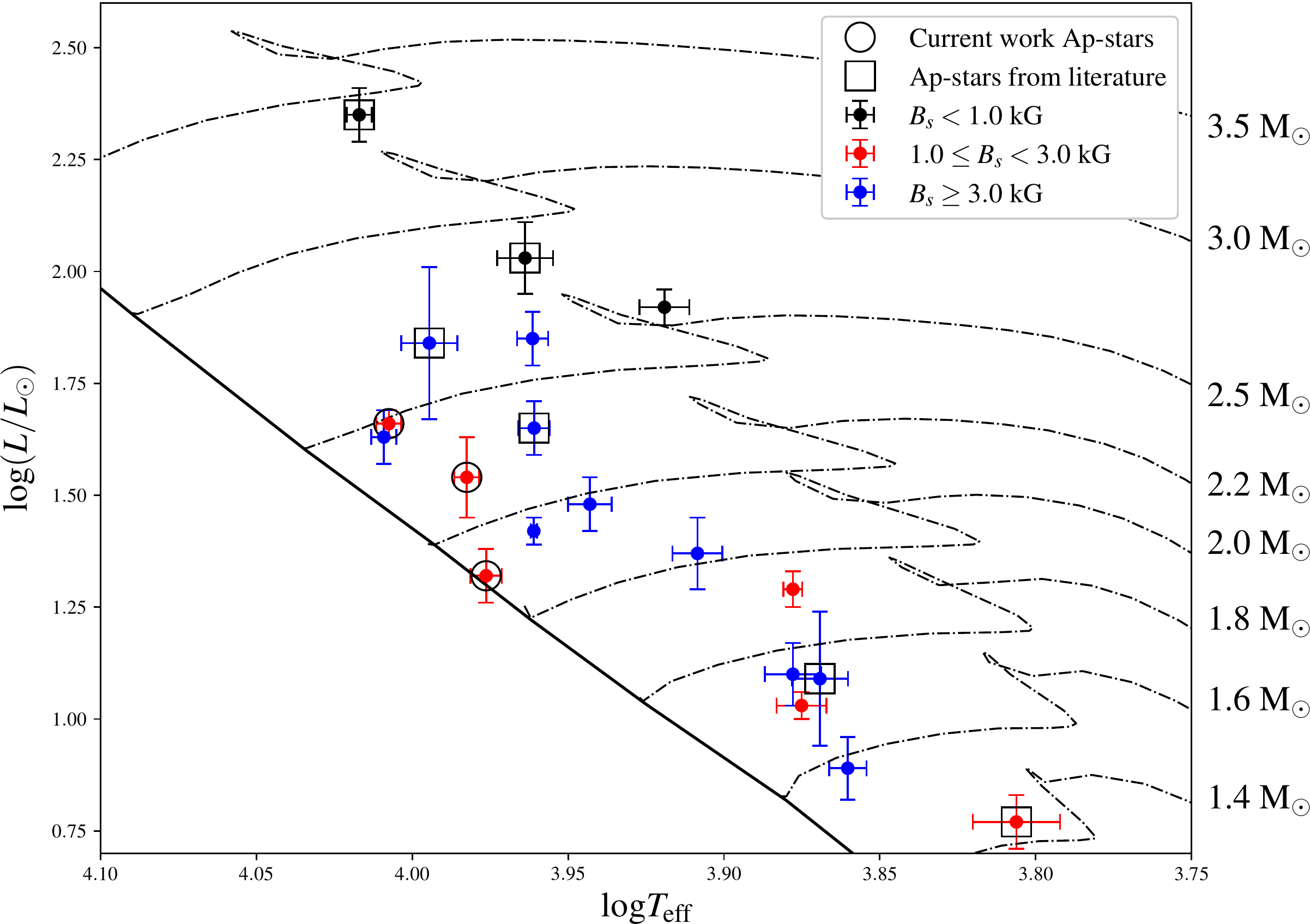}\\
         \caption{Ap stars in the HR diagram. The positions of stars from this work are marked with open circles. Six stars that were not part of the interferometric
         sample are marked with open squares. The symbols are colour-coded according to the surface magnetic field strength (see figure legend).}\label{fig:hrd}
\end{figure}

Next, we used our spectroscopically derived parameters to place stars in the Hertzsprung-Russell (HR) diagram, which we show in Fig.~\ref{fig:hrd}. Ap stars are assumed to have a global metallicity close to the solar metallicity because all the observed abundance anomalies are created in the very thin upper part of stellar envelope (atmosphere). Therefore we used the theoretical evolution tracks for solar metallicity models taken from \citet{2000A&AS..141..371G} in order to be consistent with \citet{2020A&A...642A.101P}. Solar metallicity tracks were also employed in the previous evolutionary studies of Ap stars \citep{2000ApJ...539..352H, 2006A&A...450..763K}. A trend of a decreasing strength of the magnetic field with age for stars with masses $>$2.5$M_{\odot}$ is visible. Three stars that are very close to their end on the MS life (HD~204411, HD~40312, and HD~112185 marked as black circles in the plot) have the weakest magnetic fields. However, the dependence of the magnetic field strength on age becomes less clear for stars with lower masses, but the limited sample of stars does not allow us to make other conclusions about evolutionary changes in the magnetic fields. We only note that all three stars we analysed are relatively young, but have weaker magnetic fields than the older stars in the same mass range. More work is needed to populate the HR diagram with stars that are analysed with our spectroscopic method in order to provide a consistent view of the evolution of the magnetic field in all mass ranges.

\begin{table*}[hbt!]
        \label{Fund_param_ApStars}
        \caption{Fundamental parameters of the sample of Ap stars derived with high-resolution spectroscopy.}
        \tiny
        \centering
    \begin{tabular}{r |cllc| clc | llc |c}
    \hline Stars & \multicolumn{4}{c|}{Spectroscopy} & \multicolumn{3}{c|}{Interferometry} &  \multicolumn{3}{c|}{Photometry}  &  \\
  \hline 
HD & \teff, K           &\lgg           & $R/R_{\odot}$ & log$L/L_{\odot}$ & \teff, K &  $R/R_{\odot}$ & log$L/L_{\odot}$ & \teff, K &  $R/R_{\odot}$ & log$L/L_{\odot}$  & $B_s$, kGs \\
  \hline
  4778 & ~9605(~92)$^{(a)}$ & 4.00:~~   & 2.14(4)       & 1.54(3)       & 9135(400)       & 2.36(12)      & 1.54(5)       & 9977(300)     & 1.91(4)       & 1.51(7) & 2.6 \\
 24712 & ~7250(100)$^{(b)}$ & 4.10(15)  & 1.77(4)       & 0.89(7)       & 7235(280)       & 1.75(5)       & 0.88(7)       & 7194(200)     & 1.76(4)       & 0.87(4) & 2.3 \\
108662 & 10212(100)$^{(c)}$ & 4.20:~~   & 2.09(10)      & 1.63(6)       & 8880(330)       & 2.59(12)      & 1.58(6)       &10495(400)     & 2.43(1)       & 1.81(7) & 3.3 \\
118022 & ~9140(~30)$^{(d)}$ & 4.20(2)   & 2.16(6)       & 1.47(3)       & 9100(190)       & 2.17(6)        & 1.46(5)      & 9057(290)     & 2.18(4)       & 1.46(4) & 3.0 \\
120198 & 10174(100)$^{(a)}$ & 4.20:~~   & 2.18(4)       & 1.66(3)       & 9865(370)       & 2.28(10)      & 1.65(4)       &10543(390)     & 2.01(1)       & 1.65(6) & 1.6 \\
128898 & ~7500(130)$^{(e)}$ & 4.10(15)  & 1.94(1)       & 1.03(3)       & 7420(170)       & 1.97(6)        & 1.02(2)      & 7673(190)     & 1.88(7)       & 1.04(1) & 2.0 \\
137909 & ~8100(150)$^{(f)}$ & 4.00(15)  & 2.47(7)       & 1.37(8)       & 7980(180)       & 2.63(9)        & 1.40(5)      & 7430(200)     & 3.17(9)       & 1.44(2) & 5.4 \\
153882 & ~9150(100)$^{(g)}$ & 4.01(13)  & 3.35(3)       & 1.85(3)       & 8980(600)       & 3.46(37)      & 1.85(4)       & 9727(290)     & 3.53(23)      & 2.00(11)      & 3.8 \\
176232 & ~7550(~50)$^{(h)}$ & 3.80(10)  & 2.46(6)       & 1.29(4)       & 7900(190)       & 2.21(8)        & 1.23(4)      & 7925(200)     & 2.43(1)       & 1.32(4) & 1.5 \\
188041 & ~8770(150)$^{(i)}$ & 4.20(10)  & 2.39(7)       & 1.48(3)       & 9000(360)       & 2.26(5)        & 1.48(6)      & 8433(200)     & 2.80(9)       & 1.55(7) & 3.6 \\
201601 & ~7550(150)$^{(f)}$ & 4.00(10)  & 2.07(5)       & 1.10(7)       & 7253(235)       & 2.11(7)        & 1.04(4)      & 7620(200)     & 2.04(3)       & 1.10(3) & 4.0 \\
204411 & ~8300(150)$^{(i)}$ & 3.60(10)  & 4.42(15)      & 1.92(6)       & 8520(220)       & 4.23(11)      & 1.93(5)       & 8749(300)     & 4.12(1)       & 1.95(6) & 0.8 \\
220825 & ~9470(100)$^{(a)}$ & 4.20:~~   & 1.71(4)       & 1.32(4)       & 8790(230)       & 1.78(3)        & 1.24(4)      & 9078(290)     & 1.92(3)       & 1.35(4) & 2.0 \\
\hline 
  \end{tabular}
  \tablefoot{In the \lgg\, column, the colon indicatess that the value was refined according to the hydrogen line profile.}
  \tablebib{ $^{(a)}$ Current work, $^{(b)}$ \citet{2009A&A...499..879S}, $^{(c)}$\citet{2020INASR...5..219R} , $^{(d)}$\citet{2019A&A...621A..47K} , $^{(e)}$\citet{2009A&A...499..851K}, $^{(f)}$\citet{2013A&A...551A..14S}, $^{(g)}$\citet{2020AstL...46..331R}, $^{(h)}$\citet{2013A&A...552A..28N}, $^{(i)}$\citet{2019MNRAS.488.2343R}.
  }
  \end{table*}

%--------------------------------------------------------------------

\section{Conclusions}
We derived the fundamental parameters of the three Ap stars HD~4778, HD~120198, and HD~220825, thereby completing the detailed spectroscopic
analysis of a sample of Ap stars, for each of which independent estimates of their parameters were available from interferometric observations.
A rather high rotation velocity \vs\, derived for all three stars complicated the determination of the chemical composition in their atmospheres. Therefore the abundances were obtained mainly through a spectrum synthesis method by taking the contribution of numerous line blends into account. The chemical composition of our targets shows the typical abundance pattern found in Ap stars: a deficiency of light elements, and an almost gradual increase in the overabundance towards the heavy elements, reaching +3--4~dex overabundance in the REE sequence. The only exception is Ba, whose abundance drops to the solar value. This is opposite to Ba abundance behaviour in superficially normal A and Am stars, in which overabundances of heavy elements are also observed, and Ba is the most overabundant element in the Sr-Y-Zr-Ba-Nd sequence \citep{2020MNRAS.499.3706M}. This may indicate that different mechanisms produce an overabundance  of heavy elements in the atmospheres of different groups of A stars.

We did not find any significant Fe and Cr stratification as is typical for cooler Ap atmospheres. This might be caused by a difficulty in sampling the suitable lines for stratification analysis because the rotation velocities are rather high. Therefore the final modelling was performed assuming chemically homogeneous atmospheres. 

Fundamental parameters derived by self-consistent spectroscopic modelling that took the anomalous chemical composition of stellar atmospheres  into account were compared with the parameters obtained by direct interferometric measurements. We found that stellar radii and luminosity agree within the errors, while spectroscopically determined effective temperatures are higher than those derived by interferometry for hotter stars with \teff$>$9000~K. The possible reason for this discrepancy is the underestimation of the UV contribution to the total flux balance in the interferometric studies, and/or the intrinsic inaccuracy in radius measurements for stars with small angular diameters. 
In the case of HD~4778, we succeeded in constructing a model atmosphere when only two points of Johnson's photometry were available in the optical region.

 Using HD~220825 as an example, we estimated the possible effect of the surface chemical inhomogeneity on the derived fundamental parameters through modelling the observed SED at different phases of stellar rotation. We found out  that the surface flux variability translates into $\pm$100~K effective temperature variations, which is a typical error for Ap stars and thus can be neglected in the determination of fundamental parameters in future works.

\begin{acknowledgements}
This research has made use of the data from GAIA DR2 catalogues through the VizieR catalogue access tool. The use of the VALD database is acknowledged. This work is based on observations obtained with ESpAdOnS spectropolarimeter. We use the archived data from EsPaDOnS spectrograph with programme ID: ID~09BQ78 (HD~4778), ID~16AC27 (HD~120198), and ID~18BC22 (HD~220825).
%Authors thanks T.Sitnova who provides us with NLTE departure coefficients for O\ione\ lines. 
A. Romanovskaya acknowledges the financial support from the grant RFBR, project number 19-32-90147. 
D. Shulyak acknowledges the financial support from the State Agency for Research of the Spanish MCIU through the "Center of Excellence Severo Ochoa" 
award to the Instituto de Astrof\'{\i}sica de Andaluc\'{\i}a (SEV-2017-0709).

\end{acknowledgements}

%\bibliographystyle{aa} % style aa.bst
%\bibliography{References_ALL}
%\input{aanda.bbl}
\bibliography{aanda}
\end{document}